\documentclass[a4paper, onecolumn, 11pt, accepted=2023-10-31]{quantumarticle} 
\pdfoutput = 1
\usepackage[utf8]{inputenc}
\usepackage[english]{babel}
\usepackage[T1]{fontenc}
\usepackage{amsmath, amssymb}
\usepackage{tikz}
\usepackage{lipsum}

\usepackage{url}

\usepackage{breakurl}

\usepackage{bbold}
\usepackage{float}
\usepackage{graphicx}
\usepackage{dcolumn}
\usepackage{bm}
\usepackage{braket}
\usepackage{mathtools}
\usepackage{todonotes}
\usepackage{comment}
\usepackage{microtype}
\usepackage[numbers, sort&compress]{natbib}
\usepackage{textcase}
\usepackage{hyperref} 

\begin{document}
\title{NISQ-compatible approximate quantum algorithm for unconstrained and constrained discrete optimization}

\author{M. R. Perelshtein} 
\email{mpe@terraquantum.swiss}
\affiliation{Terra Quantum AG, Kornhausstrasse 25, 9000 St. Gallen, Switzerland}
\affiliation{QTF Centre of Excellence, Department of Applied Physics, Aalto University, P.O. Box 15100, FI-00076 AALTO, Finland}
\affiliation{InstituteQ – the Finnish Quantum Institute, Aalto University, Finland}

\author{A. I. Pakhomchik}
\affiliation{Terra Quantum AG, Kornhausstrasse 25, 9000 St. Gallen, Switzerland}

\author{Ar.~A.~Melnikov}
\affiliation{Terra Quantum AG, Kornhausstrasse 25, 9000 St. Gallen, Switzerland}

\author{M.~Podobrii}
\affiliation{Terra Quantum AG, Kornhausstrasse 25, 9000 St. Gallen, Switzerland}

\author{A.~Termanova}
\affiliation{Terra Quantum AG, Kornhausstrasse 25, 9000 St. Gallen, Switzerland}

\author{I.~Kreidich}
\affiliation{Terra Quantum AG, Kornhausstrasse 25, 9000 St. Gallen, Switzerland}

\author{B.~Nuriev}
\affiliation{Terra Quantum AG, Kornhausstrasse 25, 9000 St. Gallen, Switzerland}

\author{S.~Iudin}
\affiliation{Terra Quantum AG, Kornhausstrasse 25, 9000 St. Gallen, Switzerland}

\author{C.~W.~Mansell}
\affiliation{Terra Quantum AG, Kornhausstrasse 25, 9000 St. Gallen, Switzerland}

\author{V.~M.~Vinokur}
\affiliation{Terra Quantum AG, Kornhausstrasse 25, 9000 St. Gallen, Switzerland}
\affiliation{Physics Department, City College of the City University of New York, 160 Convent Ave, New York, NY 10031, USA}

\begin{abstract}
\noindent
Quantum algorithms are gaining extreme popularity due to their potential to significantly outperform classical algorithms. Yet, practical applications of quantum algorithms to optimization problems meet challenges related to the efficiency of the existing quantum algorithms training, the shape of their cost landscape, the accuracy of their output, and their ability to scale to large-size problems.
Here, we present a 
gradient-based quantum algorithm for hardware-efficient circuits with amplitude encoding.
We show that simple linear constraints can be directly incorporated into the circuit without additional modification of the objective function with penalty terms.
We employ numerical simulations to test it on {\tt MaxCut} problems with complete weighted graphs with thousands of nodes and run the algorithm on a superconducting quantum processor.
We find that being applied to unconstrained {\tt MaxCut} problems with more than 1000 nodes, the hybrid approach combining our algorithm with a classical solver called CPLEX realizes a better solution than the CPLEX alone. 
This demonstrates that hybrid optimization is one of the leading use cases for modern quantum devices.
    
\end{abstract}

\maketitle

\section{Introduction}\label{sec:Introduction}

Recent years marked an experimental breakthrough demonstrating that quantum computers can outperform classical supercomputers in solving certain complex problems\,\cite{AruteAndOthers2020, PhysRevLett.127.180501AndOthers, ZHU2022240AndOthers}.
Yet, since quantum computers are now at a very early stage of development, their optimal use is achieved 
via combining them with the high-performance classical approaches\,\cite{doi:10.7566/JPSJ.90.032001}. One of the focuses of this direction is using for this hybridization the Noisy Intermediate-Scale Quantum (NISQ) devices. 
As we have shown in Ref.\,\cite{TQwhitepaper}, such a hybridization\,\cite{bravyi2016trading, McClean_2016, li2017hybrid, zhu2019trainingAndOthers} is a very promising development. 
Hybrid pipelines that combine both quantum and classical approaches are especially beneficial for solving complex optimization problems~\cite{ajagekar2020quantum,shaydulin2019hybrid}.
While numerical optimization remains one of the most challenging and demanding problems, hybrid algorithms for {\tt MaxCut} problem\,\cite{caha2022twisted}, Knapsack~\cite{haddar2016hybrid}, Unit Commitment~\cite{mahroo2022hybrid}, and general binary optimization approaches\,\cite{tran2016hybrid, liu2018green} have already been put forth. 

Variational quantum algorithms (VQAs), where the parameterized quantum network is iteratively optimized via classical computing~\cite{CerezoAndOthers2021, mugel2021hybrid, ge2022optimization} are one of the most promising among these hybrid quantum-classical methods. 
Finding the best ways to efficiently train these algorithms to produce an accurate output is an outstanding challenge.
One of the first proposed VQAs was the quantum approximate optimization algorithm (QAOA)\,\cite{QAOA} for solving combinatorial optimization problems, in particular, a problem of discrete optimization known as a quadratic unconstrained binary optimization (QUBO)\,\cite{Willsch2020, Lykov2022}. 
This problem is of high interest due to its relationship with other optimization problems, such as, for example, portfolio optimization\,\cite{venturelli2019reverse}, the factorization of large integers\,\cite{peng2019factoring}, and other graphs-related problems\,\cite{glover2018tutorial}.
As is the case with many algorithms designed for the NISQ devices, recent experiments highlighted challenges in implementing the QAOA on the problem graphs the configuration of which differs from the native hardware topology, even for small system sizes\,\cite{AruteAndOthers2020}. 
Consequently, it is important to explore hardware-eﬃcient approaches using a series of native approaches to the quantum platform gates whose connectivity is decoupled from the topology of the QUBO problem itself\,\cite{CerezoAndOthers2021, benedetti2021hardware}.

A further issue of many NISQ algorithms, such as the VQAs and quantum annealing\,\cite{yarkoni2021}, is that the number of qubits is roughly equal to the number of classical variables, making it hard to solve problems at practically interesting scales.
Recently, a hardware-efficient method of encoding $n_c$ classical variables into the amplitudes of ${\cal O}$($\log{n_c}$) qubits was proposed in Ref.\,\cite{Tan2020} for the sparse QUBO problems.
The variational quantum state represents a probability distribution over all the classical solutions and the learning process aims at producing a state with a maximal probability associated with the optimal solutions.
However, the question of how to efficiently train such circuits is an open question, and gradient-free optimization methods may limit the problem size\,\cite{Liu2018Differentiable}. 

General combinatorial optimization problems are either constrained or unconstrained, depending on what restrictions are imposed on the variables.
The usual approach, which is also popular in quantum annealing, is to add a penalty term to the problem objective function. 
This converts a constrained optimization problem into an unconstrained one where minimizing the cost function leads to the constraints being satisfied. 
This approach requires careful tuning of the penalty terms and leads to a complex search space where the cost landscape is plagued with local minima, complicating the algorithm's convergence.
Alternative methods for including constraints have been investigated~\cite{matsuo2020problem, gilliam2021grover, Niroula2022}, however, a scalable approach was not proposed yet.

Here, addressing the training and constraints issues, we present a gradient-based algorithm for the hardware-efficient quantum circuits with the amplitude encoding, \textcolor{black}{see Sec.~\ref{sec:UnconstrainedOptimization}}.  
We study this Quantum Encoding ({\it QuEnc}) method in\,\cite{QuEnc_Patent, QuEnc_Patent2} and find that simple linear constraints can be directly incorporated into the circuit without additional modification of the objective function with penalty terms, \textcolor{black}{see Sec.~\ref{sec:constraints}}.
We test the algorithm on {\tt MaxCut} problems~\cite{Karp1972} with complete weighted graphs consisting of thousands of nodes, \textcolor{black}{see Sec.~\ref{sec:Results}}.
We conduct a thorough analysis, \textcolor{black}{see Sec.~\ref{sec:Analysis}}, of the algorithm's ability to perform a global optimum search through numerical calculations using QMware~\cite{QMWare} and via experiments using superconducting devices from IBMQ~\cite{IBMQ}, \textcolor{black}{see Sec.~\ref{sec:Methods}}. 
We use this quantum algorithm in combination with state-of-the-art classical solvers and show that for unconstrained {\tt MaxCut} problems with more than 1000 nodes, the hybrid approach can find a better solution than the classical solver (CPLEX) alone.
This demonstrates that quantum algorithms can serve as a booster for classical solvers and help them find better solutions in less time.

\section{Unconstrained Optimization}\label{sec:UnconstrainedOptimization}

In this section, we describe the QUBO problems, \textcolor{black}{see Sec.\,\ref{subsec:TheQUBOproblem}}, and show how they are encoded in quantum circuits, \textcolor{black}{see Sec.~\ref{subsec:QuantumCircuitForTheQUBO}}.
We then discuss how these circuits learn, \textcolor{black}{see Sec.\,\ref{sec:Learning}}, and how they can be combined with classical methods, \textcolor{black}{see Sec.\,\ref{subsec:HybridizationOfQuantumAlgorithm}}. 
\textcolor{black}{The implementation details are described in Sec.\,\ref{sec:Methods} and the results are presented in Sec.\,\ref{sec:Results}.}

\subsection{The QUBO problem}\label{subsec:TheQUBOproblem}

\begin{figure}[ht]
    \centering
    \includegraphics[width=55mm]{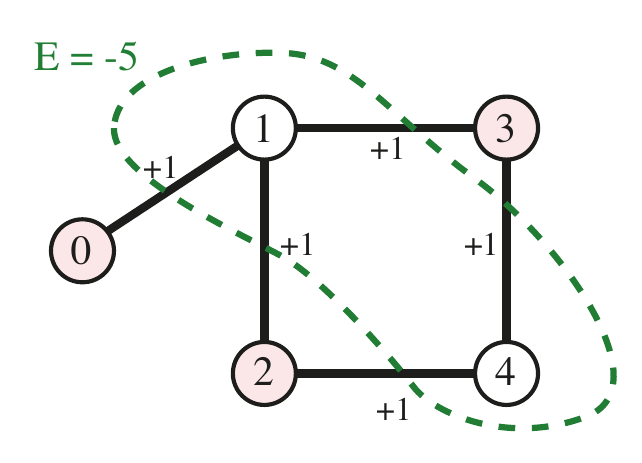}
    \caption{
    The green dashed curve represents the maximum cut of the depicted graph, since it includes all edges. Different vertex colors indicate different subgroups of vertices. According to Eq.~\eqref{energy}, the energy value for this cut is $-5$.
    }
    \label{fig:MaxCut}
\end{figure}

Quantum discrete optimization emerged in the context of the ground state search problem in the scope of the Ising model~\cite{santoro2006optimization}.
This model is equivalent to the QUBO which generally is an NP-hard discrete optimization problem where a cost function, $C$, which is a quadratic function of classical binary variables, is minimized. 
That is, the task is to find $\min_{\vec{x}}(C)=\min_{\vec{x}}(\vec{x}^TQ\vec{x})$, where $\vec{x} \in \{0,1\}^{n_c}$ is a vector of the $n_c$ binary variables that correspond to the $n_c$ nodes in the problem graph, see Fig.~\ref{fig:MaxCut}, and $Q$ is a real, upper triangular matrix corresponding to the weights of the graph's edges.

One of the canonical and well-studied QUBO problems is the {\tt MaxCut} problem, which is a search for the partition of the graph's nodes into two complementary sets, such that the total weights of edges between these two sets is as large as possible.
A node is a member of one set if its value is zero and a member of the other set if its value is one.
An example of finding the maximum cut of an unweighted 5-node graph is shown in Fig.~\ref{fig:MaxCut}.
The QUBO version of the arbitrary {\tt MaxCut} problem is based on a minimization of the following quadratic function
\begin{equation}
    E = -\sum_{i,j=0}^{n_c-1}d_{ij}(x_i-x_j)^2,
    \label{energy}
\end{equation}
where $d_{ij}$ is the weight of the edge between $i$th and $j$th nodes in the studied graph.
The solution to this minimization problem is a binary string $\vec{x}$ of the nodes' indicators, which show the correspondence to one of two sets.
The elements of the QUBO matrix, in turn, are $Q_{ij}=2d_{ij} (i > j)$ and $Q_{ii}=-\sum_j d_{ij}$.
Therefore, the cost function to optimize is $C=\vec{x}^T\,Q\,\vec{x}$.

The Hamiltonian of the Ising model is
\begin{equation}
    H = \sum_{i \in I} h_i s_i + \sum_{i \neq j} J_{ij} s_i s_j,
    \label{Ising}
\end{equation}
where $s_i \in \{-1, 1\}$ is a spin, $h_i \in \mathbb{R}$ is the external magnetic field in the vicinity of spin $i$, $I$ is the set of all the spins, and $J_{ij} \in \mathbb{R}$ is the coupling between spin $i$ and spin $j$.
Generally, it is not equivalent to the {\tt MaxCut} model due to the presence of the linear term.
However, 
as shown in Appendix~\ref{AppendixSecProof}, it can be reduced to {\tt MaxCut} by adding a single ancilla qubit $a$, which has to be connected to all qubits with the coupling strength $J_{ia} = h_i$.
This leaves only quadratic terms in the Hamiltonian making it equivalent to the well-known spin glass Hamiltonian~\cite{MaxCutIsingGlasses, santoro2002theory}
\begin{equation}
    H_g = \sum_{i \in I} h_i s_i s_a + \sum_{i \neq j} J_{ij} s_i s_j = \sum_{i \neq j, i, j \in I \bigcup \{a\}} J^{g}_{ij} s_i s_j.
    \label{Hg}
\end{equation}
It is possible to convert spins to bits via the simple transformation $x_i = (s_{i} + 1)/2$.

\begin{figure*}
    \noindent\centering{
    \includegraphics[width=0.9\textwidth]{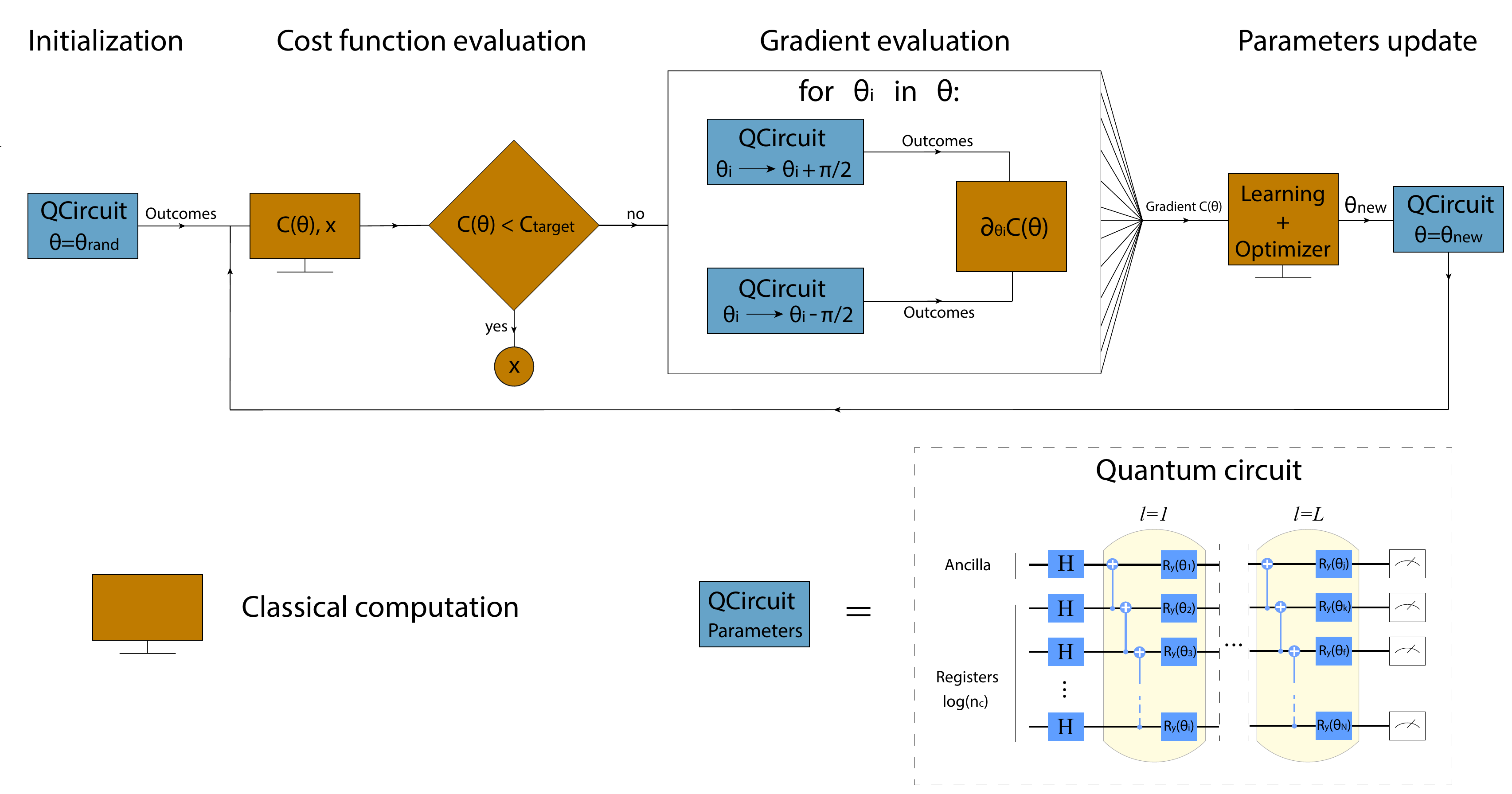}} 
    \caption{
    Hybrid quantum classical computational network based on the descent minimization of the quantum gradient.
    The workflow is as follows.
    Initialization: we randomly initialize the parameters vector $\vec{\theta}$.
    Cost function evaluation: we evaluate the cost function of the solution from the circuit outcomes using a classical computer and check the convergence condition.
    If this condition is met, we abort the algorithm.
    If not, we proceed to the next stage. 
    Gradient evaluation: we extract the partial derivative using the quantum computational network (QCN), accumulate the derivatives in order to extract the gradient and then perform the 
    descent procedure with additional potential optimization, e.g., ADAM optimization.
    Parameters update: we update $\vec{\theta}$ and calculate the cost function from the outcomes of the circuit with updated parameters.
    We iteratively perform this procedure until the convergence condition is met.
    Quantum circuit: we refer to this particular quantum circuit as to the Sequential-2QG ansatz.
    } 
    \label{qcn}
\end{figure*}

\subsection{Quantum circuit for the QUBO}\label{subsec:QuantumCircuitForTheQUBO}

While most variational quantum algorithms~\cite{CerezoAndOthers2021} utilize a complete encoding where each bit from the $n_c$ variables is represented by a single qubit, Ref.~\cite{Tan2020} proposed an amplitude encoding method.
It only needs a logarithmic number of register qubits plus some additional ancilla qubits to capture the correlations between the variables.
In the simplest case of a single ancilla, the problem is reduced to minimal encoding.

The quantum circuit for the minimal encoding scheme consisting of $\log{n_c}$ register qubits, one ancilla qubit, and $L$ layers of gates is presented in Fig.~\ref{qcn}. 
While a wide variety of gate layouts are possible, we design one where each layer consists of both variational single-qubit gates, $R_y(\theta)$, and the CNOT gates that connect all qubits in the array.
This circuit is our ``Sequential-2QG'' ansatz.
\textcolor{black}{The expressibility of this ansatz is defined in Appendix~\ref{AppendixSecExpressibility} and analyzed in Sec.~\ref{subsec:AnsatzAnalysis}.}

The parameterized state of the variational circuit in the minimal encoding scheme is given by 
\begin{equation}
    \ket{\psi(\vec{\theta})} = \sum_{i=0}^{n_c-1} \beta_i(\vec{\theta})(a_i(\vec{\theta})\ket{0}_a+b_i(\vec{\theta})\ket{1}_a) \otimes \ket{i}_r,
    \label{eq:quenc_quantum_state}
\end{equation} 
where the subscript $a$ represents the single ancilla qubit and subscript $r$ represents the $\log{n_c}$ register qubits.
Every $\ket{i}_r$ state, a product of computational basis states, being interpreted as a binary number, denotes one of the $n_c$ nodes of the problem graph.
For each term in the overall superposition, the singe-qubit ancilla state indicates whether its node-associated register state should be zero or one. 
That is, by measuring the parameterized state in the computational basis, we determine that the $i^{th}$ classical variable is a $0$ with probability Pr$(x_i=0)=|a_i(\vec{\theta})|^2$ and a $1$ with probability Pr$(x_i=1)=|b_i(\vec{\theta})|^2$. \textcolor{black}{The result of the optimization is a binary vector $\vec{x}$, which has maximal probability, i.e. $x_i=1$ if $|b_i(\vec{\theta})|^2>|a_i(\vec{\theta})|^2$, otherwise $x_i=0$.}

Our goal is to choose $L\times(\log(n_c)+1)$-dimensional vector, $\vec{\theta}$ in a way that the sampling of the final state $\ket{\psi(\vec{\theta})}$ \textcolor{black}{gives} the optimum binary solution.
To that end, we need to minimize a cost function, which contains the information about the problem in the form of the QUBO matrix $Q$%
\begin{equation}
    \mathcal{C} = \sum_{i \neq j} Q_{ij} |b_i(\vec{\theta})|^2 |b_j(\vec{\theta})|^2 + \sum_{i = 0}^{n_c-1} Q_{ii} |b_i(\vec{\theta})|^2.
    \label{cost}
\end{equation}
Here, the classical cost function $C=x^TQx$ is replaced with the function, $\mathcal{C}$ of continuous quantum amplitudes such that $\min{\mathcal{C}}=\min{C}$~\cite{Tan2020}.

Essentially, this method allows us to solve a discrete problem as a continuous-variable optimization using exponentially fewer classical resources.
A clear bottleneck is the training of such a model: the gradient, just like the objective function itself, tends to be classically intractable because increases in the number of qubits can eliminate all profits gained from the quantum approach.

One approach consists in using gradient-free methods~\cite{nesterov2017random}, e.g., COBYLA~\cite{powell2007view}. 
However, these methods are prone to early convergence in local optima when the problem is multimodal and the starting point is far from the optimum~\cite{powell2007view}.
As a consequence, the final solutions are not as precise as solutions given by gradient-based methods, and the number of quantum circuit calls may be too large for being convergent.
Fortunately, quantum computing allows us to calculate gradients using quantum networks with the same level of complexity as an objective circuit~\cite{Schuld2019}, \textcolor{black}{
enabling an alternative, gradient-based method for optimization. 
Due to its better performance relative to COBYLA in some preliminary numerical comparisons, and due to its ability to compute all the partial derivatives in parallel, we chose to employ this method in this work. 
See the following subsection (Sec.~\ref{sec:Learning}) for further details.}

\subsection{Learning of the quantum circuit}
\label{sec:Learning}

In 2019, Schuld {\it et al.} introduced the parameters-shift rule to calculate the gradient of a gate in the form of $e^{-i\mu G}$ via evaluation of the original expectation twice, but with one circuit parameter shifted by a fixed value~\cite{Schuld2019}.
Here, $G$ is the Hermitian generator with at most two distinct eigenvalues, and $\mu$ is the gate parameter to be optimized. 

The studied quantum circuit is represented as a function that maps the $N$-dimensional vector of gate parameters $\vec{\theta}$ to an expectation value $f(\vec{\theta})$ $=\bra{\psi{(\vec{\theta})}} \hat{A} \ket{\psi{(\vec{\theta})}}$ of the measurement of some observable $\hat{A}$. 
In our case, $N=L\times(\log(n_c)+1)$.
In order to calculate the gradients of a cost function that takes the expectation as an argument, we evaluate the partial derivatives $\partial_{\theta_j}f(\vec{\theta})$, where $j \in [1, N]$.
We calculate the derivative of $i^{th}$ single-qubit $R_y(\theta_j)$ gate by sampling the original circuit, where this gate is replaced by $R_y(\theta_j\pm\pi/2)$.
The expectation values of these circuits are $f([\theta_1, \cdots, \theta_j\pm\pi/2, \cdots, \theta_N]) \equiv f(\theta_j\pm\pi/2)$ and the partial derivative is given by 
\begin{equation}
    \partial_{\theta_j}f(\vec{\theta})=\frac{1}{2}\left(f(\theta_j+\pi/2)-f(\theta_j-\pi/2)\right).
    \label{paramshift}
\end{equation}

We use this formula to calculate the derivatives of the expectations of the studied observables as well as the cost function's derivative, $\partial_{\theta_j}\mathcal{C}$.
In the latter case, we express the derivative of $|b_i(\vec{\theta})|^2$ through expectations and leverage the chain rule,
see Appendix~\ref{AppendixSecDerivative},
%
\begin{align}
    \label{grad_calc}
    \partial_{\theta_j} \mathcal{C} &= \sum_{i \neq k} Q_{ik} \left( \partial_{\theta_j} |b_i(\vec{\theta})|^2 |b_k(\vec{\theta})|^2 
    + 
    |b_i(\vec{\theta})|^2
    \partial_{\theta_j}
    |b_k(\vec{\theta})|^2
    \right) 
    + \sum_{i=0}^{n_c-1} Q_{ii} \partial_{\theta_j} |b_i(\vec{\theta})|^2.
\end{align}

For the learning procedure, after the random initialization of $\vec{\theta}=\vec{\theta}_0$ and the evaluation of the initial cost function $\mathcal{C}_0$, we calculate the gradient $\nabla \mathcal{C}|_{\vec{\theta}=\vec{\theta}_0}$ and update the parameters using the gradient descent
\begin{equation}
    \vec{\theta}_{t+1} = \vec{\theta}_{t} - \alpha \nabla \mathcal{C}|_{\vec{\theta}=\vec{\theta}_{t}},
    \label{grad_dec}
\end{equation}
where $\alpha$ is a learning rate hyperparameter and $t$ is an iteration index.
Overall, each step of gradient descent requires two experiments for each partial derivative and a single experiment for the cost function evaluation.


Just as gradient methods can encounter issues in classical neural networks, gradient descent in quantum circuits can sometimes struggle to provide stable performance without some additional optimization.
Since we use the probability distributions that are provided by the quantum processor for the gradient calculations, we inevitably face noisy data.
It is important to note that various optimization techniques inherited from classical deep learning (e.g., ADAM \textcolor{black}{(see Ref.~\cite{Kingma2015AdamAM} and Appendix~\ref{AppendixSecADAM})}, stochastic gradient descent and cost function regularization) can be used to improve the network performance.

We illustrate the whole computational network, both classical (orange) and quantum (blue) parts, in Fig.~\ref{qcn}. 
First, we initialize the circuit, which can be done either randomly or via using a predefined state (warm-start).
We evaluate the cost function Eq.~\eqref{cost} by sampling the state and checking the satisfactory requirement -- in the case where the objective value is not sufficiently small, we calculate gradients using Eq. ~\eqref{grad_calc} and update the circuit's parameters $\vec{\theta}$ using Eq.~\eqref{grad_dec} with classical optimizers if needed. 
When we aim to obtain a solution with the known target cost function $\mathcal{C}_{\mbox{target}}$, we update the parameters of our circuit until this value is reached.
In the situation where the target energy is unknown, the gradient descent procedure is performed until the convergence.

\subsection{Hybridization of quantum algorithm}\label{subsec:HybridizationOfQuantumAlgorithm}

A variational quantum algorithm is a form of hybrid quantum computing where a quantum circuit is optimized using classical methods.
In our case, we calculate the objective function using a classical computer, but the gradient calculation is performed using quantum circuits.

Our QuEnc algorithm has warm-start capabilities,
see Appendix~\ref{AppendixSecWarm}.
That is, an initial or preliminary solution can be used as a guess for QuEnc, which is then improved.
The result of the QuEnc optimization can also serve as a booster for the classical optimizer that utilizes the obtained solution as a starting point.
Therefore, the quantum algorithm is considered as a two-port (input, output) element in the optimization scheme that can be utilized according to the problem in order to maximize the performance of the full optimization pipeline. 
The optimal strategy for the use of QuEnc in such pipelines is highly dependent on the problem.
Here, in order to showcase the capabilities of the hybrid solution, we realize the simplest insertion of the QuEnc solution into the classical solver as an initial guess.

\section{Constrained optimization}
\label{sec:constraints}

So far, we have analyzed only a single objective function with binary variables that is minimized by the tuning and sampling of a variational circuit in order to produce bitstrings. 
In constrained optimization, some bitstrings may violate the constraints making them unfeasible solutions.
While adding a penalty term usually fails to provide a feasible bitstring, a more rigorous approach that restricts the search space of the solutions to be inside the constrained space is required.
Here, addressing this issue, we propose techniques to encode simple linear constraints directly into QuEnc's quantum circuit.
\textcolor{black}{The implementation details are described in Sec.~\ref{sec:Methods} and the results are presented in Sec.~\ref{sec:Results}.}

\subsection{Two variables with linear constraint}
\label{simplest_constraint}
From the illustrative perspective, let us consider a toy-constrained optimization problem with two variables $x_0$ and $x_1$:
    \begin{equation}
        \begin{aligned}
        \min( \vec{x}^T Q \vec{x})\,, 
        \\
        \text{such that: } x_0 + x_1 = 1\,.
        \end{aligned}
    \end{equation}
The obvious solution to this problem is $\vec{x} = \left( 0, 1 \right)$ or $\left( 1, 0 \right)$.
A variational circuit returns the quantum state $\ket{\psi(\vec{\theta})}$, which which corresponds to the probability distribution $\vec{P}^1(\vec{\theta})$.
The probabilistic state describes the probability of bits to take a value one $P^1_k = P(x_k = 1)$.
We evaluate the initial cost function and perform a single iteration of gradient descent.
Since we aim at constructing such a state for which $\vec{P^1}$ always satisfies the constraint (the most probable value of the bits satisfies the equation), we build the circuit that allows us to collapse the state into the space of the feasible solutions during the measurement.
Let us define the feasible subspace and its orthogonal complement space.

Two states that satisfy the constraint are
\begin{equation}
\begin{aligned}
    \ket{\psi_0} = \frac{1}{\sqrt{2}} (\ket{0}_a\ket{0}_r + \ket{1}_a\ket{1}_r), \\
    \ket{\psi_1} = \frac{1}{\sqrt{2}} (\ket{0}_a\ket{1}_r + \ket{1}_a\ket{0}_r).
\end{aligned}
\end{equation}
Note that $\alpha\ket{\psi_0}+\beta\ket{\psi_1}=\frac{1}{\sqrt{2}}(\alpha\ket{0}_a\ket{0}_r+\beta\ket{1}_a\ket{0}_r+\beta\ket{0}_a\ket{1}_r+\alpha\ket{1}_a\ket{1}_r)$
corresponds to solution (0, 1) when $|\alpha|>|\beta|$, and to (1, 0) when $|\alpha|<|\beta|$.
That is, except for the case when $|\alpha|=|\beta|$ (which has zero probability), it always corresponds to a feasible solution. 
Therefore, $\ket{\psi_0}, \ket{\psi_1}$ is a basis in the feasible subspace, \textcolor{black}{$V$}, and we can consider the corresponding operator $P = \ket{\psi_0} \bra{\psi_0} + \ket{\psi_1} \bra{\psi_1}$ that projects quantum states onto this subspace. 

The additional two states, which cover the whole computational space, are
\begin{equation}
    \begin{aligned}
        \ket{\psi_0^T} &= \frac{1}{\sqrt{2}} (\ket{0}_a\ket{0}_r - \ket{1}_a\ket{1}_r), \\
        \ket{\psi_1^T} &= \frac{1}{\sqrt{2}} (\ket{0}_a\ket{1}_r - \ket{1}_a\ket{0}_r)\,.
    \end{aligned}
\end{equation}
The corresponding projection operator $P^*=\ket{\psi_0^T} \bra{\psi_0^T} + \ket{\psi_1^T} \bra{\psi_1^T}$ represents the \textcolor{black}{orthogonal complement $V^*$ to the feasible subspace.} 
Note that all these states are the Bell states:
$\ket{\Phi^+}, \ket{\Psi^+}$ form a feasible subspace and by adding $\ket{\Phi^-}, \ket{\Psi^-}$ we cover the whole computational space.
Note further that \textcolor{black}{$V^*$ may also contain feasible states.
In the presented case, all the states in $V^*$ also satisfy the constraint, while all the unfeasible states are linear combinations of states from $V$ and $V^*$. 
This does not pose an issue since our goal is to ensure that the states from $V$ are feasible. 
However, not all the feasible states are in $V$ since, generally, the feasible states do not form a linear subspace.}

Leveraging an additional ancillary qubit in the circuit, -- which we call the constraint ancilla hereinafter -- we use the joint quantum state
$\ket{0}_C \otimes \ket{\psi(\vec{\theta})}$ in order to collapse to the feasible subspace using the following transformation:
\begin{align}
    \Big[ \mathcal{I} \otimes P + X \otimes P^* \Big]\,  (\ket{0}_C \otimes \ket{\psi(\vec{\theta})}),
    \label{eq:measure_gate}
\end{align}
where $\mathcal{I}$ is the identity gate and $X$ is the Pauli gate.
By performing a measurement on the constraint ancilla and postselecting outcomes with $0$ output, we make the state collapse to the feasible subspace.
\textcolor{black}{The probability of collapsing to feasible subspace is equal to $p_{feas}=|P\ket{\psi}|^2=|\braket{\Phi^+|\psi}\ket{\Phi^+}+\braket{\Psi^+|\psi}\ket{\Psi^+}|^2=|\braket{\Phi^+|\psi}|^2+|\braket{\Psi^+|\psi}|^2$. 
It depends on $\ket{\psi}$, so, theoretically, we can only estimate the average value of it. 
Bell states form an orthonormal basis, therefore $|\braket{\Phi^+|\psi}|^2+|\braket{\Psi^+|\psi}|^2+|\braket{\Phi^-|\psi}|^2+|\braket{\Psi^-|\psi}|^2=|\ket{\psi}|^2=1$. 
Assuming that $\ket{\psi}$ is a random vector, $\left< |\braket{\Phi^+|\psi}|^2\right>=\left<|\braket{\Psi^+|\psi}|^2\right>=\left<|\braket{\Phi^-|\psi}|^2\right>=\left<|\braket{\Psi^-|\psi}|^2\right>=1/4$ and $\left<p_{feas}\right>=1/2$. }

\subsection{Encode constraints via single- and two-qubit gates}

\begin{figure}[ht]
     \noindent\centering{
    \includegraphics[scale=0.6]{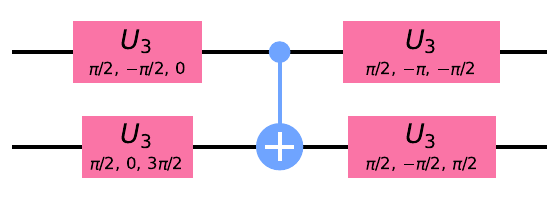}
    }
    \caption{Matrix $A$ [Eq.~\ref{eq:MatrixA}] is decomposed into CNOT and single-qubit gates. 
    An alternative decomposition would be a Hadamard gate followed by a CNOT but, without the loss of generality, the presented decomposition was derived from a general two-qubit gate.
  }
    \label{fig:A_matrix}
\end{figure}

To implement our constraints in practice, we build said projectors using single- and two-qubit gates.
First, we build a matrix $A$, which transforms the basic states $\ket{\psi_0}, \ket{\psi_1}, \ket{\psi_0^T}, \ket{\psi_1^T}$ into computational basis
\begin{align}
\label{eq:MatrixA}
    A = \frac{1}{\sqrt{2}} \begin{pmatrix}
        1 & 0 & 0 & 1 \\
        0 & 1 & 1 & 0 \\
        1 & 0 & 0 & -1 \\
        0 & 1 & -1 & 0 \\
    \end{pmatrix}.
\end{align}
This matrix is decomposed into single-qubit gates and CNOTs as demonstrated in Fig.\,\ref{fig:A_matrix}.
Applying $A$ to the basis states of $V \oplus V^*$ we obtain
\begin{equation}
\begin{aligned}
    \ket{\psi_0} \rightarrow \ket{0}_r\ket{0}_a, \ket{\psi_1} \rightarrow \ket{0}_r\ket{1}_a \\
    \ket{\psi_0^T} \rightarrow \ket{1}_r\ket{0}_a, \ket{\psi_1^T} \rightarrow \ket{1}_r\ket{1}_a 
\end{aligned}
\end{equation}
This transformation enables us to distinguish states from the $V$ and $V^*$ spaces using the state of the 
register qubit.
%

We apply the CNOT gate between the constraint ancilla $\ket{0}_C$ (target) and the register qubit (control), making the constraint ancilla an indicator of the feasible subspace.
After that, we bring the states of the qubits to their original states by applying the $A^{-1}$ operator.
The feasible $\ket{\psi}\in V$ and unfeasible $\ket{\psi^*}\in V^*$ states evolve in the following way
\begin{equation}
    \begin{aligned}
        \ket{0}_C \ket{\psi} \xrightarrow[]{A} \ket{0}_C \ket{0}_r \ket{\xi}_a \xrightarrow[]{CNOT_{r,\,C}} \\ \ket{0}_C \ket{0}_r \ket{\xi}_a \xrightarrow[]{A^{-1}} \ket{0}_C \ket{\psi}, \\[20pt]
        \ket{0}_C \ket{\psi^*} \xrightarrow[]{A} \ket{0}_C \ket{1}_r \ket{\xi^*}_a \xrightarrow[]{CNOT_{r,\,C}} \\ \ket{1}_C \ket{1}_r \ket{\xi^*}_a \xrightarrow[]{A^{-1}} \ket{1}_C \ket{\psi^*},
    \end{aligned}
\end{equation}
where $\ket{\xi}_a, \ket{\xi^*}_a$ are the unknown states of the 
ancilla qubit.
Measuring the constraint ancilla in the $0$ state collapses the wave function onto the feasible subspace $V$.

\begin{figure}[ht]
     \noindent\centering{
    \includegraphics[scale=0.35]{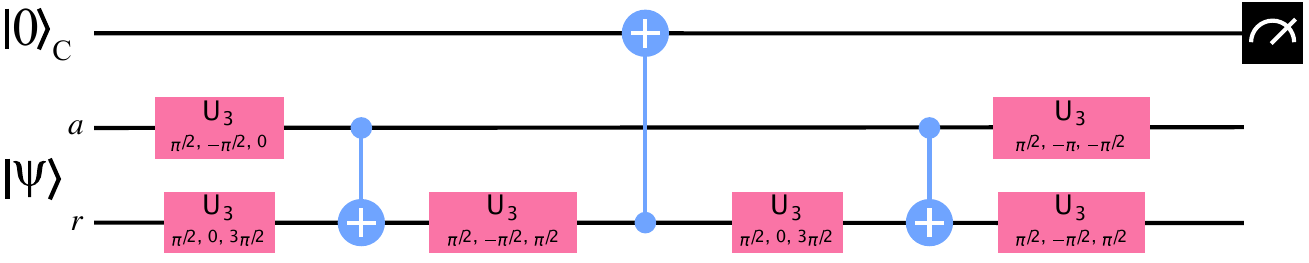}
    }
    \caption{Full circuit for separating states $V$ and $V^*$ in the case where there is only one constraint.
    }
    \label{fig:meas_full_gate}
\end{figure}

The full constraint realization for the case of two classical variables (single constraint ancilla, single algorithm ancilla, and single register qubit) is shown in Fig.~\ref{fig:meas_full_gate}.

\subsection{The arbitrary QUBO with only one constraint}
\label{sec:qubo_one_cons}

We consider an arbitrary QUBO matrix $Q\in\mathbb{R}^{N\times N}$ with a constraint $x_i + x_j = 1$ for arbitrary variables with indices $ i, j\in\{0,1,\dots,N - 1\}$.
For a {\tt MaxCut} problem, this corresponds to requiring the $i^{th}$ and $j^{th}$ nodes to be in different subsets.
We show how to reduce this problem to the case described in the previous sections by applying a transformation $(i, j) \rightarrow (0, 1)$. 

\begin{figure}[ht]
    \noindent\centering{
    \includegraphics[scale=0.5]{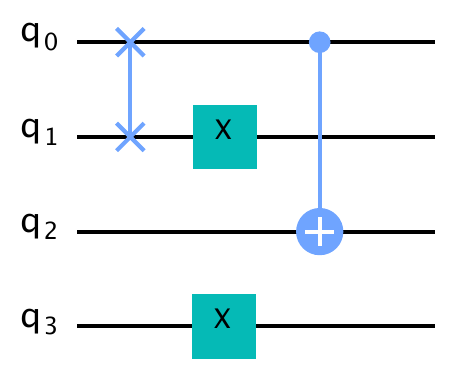}
    }
    \caption{Circuit diagram, consisting of four register qubits, for turning the constraint $x_9 + x_{15} = 1$ into the constraint $x_0 + x_1 = 1$. See main text for more details.}
    \label{fig:constraint_moving}
\end{figure}

We illustrate this transformation with the example shown in Fig.\,\ref{fig:constraint_moving}, where $i = 9$, $j = 15$, and the register qubits are $q_0$ to $q_3$.
Recall that the register qubits are part of the quantum state given by Eq.~\eqref{eq:quenc_quantum_state}.
We use the following algorithm to manipulate the terms of this superposition:
\begin{enumerate}
    \item Find the first {\it different bit} in the binary representation of the numbers $i$ and $j$.
    In the example, $i = 10{\bf0}1_2$ and $j = 11{\bf1}1_2$, so enumerating from right to left, the first {\it different} bit is the first bit (since the zeroth bit is $1$ for both numbers). 
    \item Apply a SWAP operation between the first {\it different} bit and the zeroth one.
    That is,
    $(i=10{\bf 01}_2$, $j=11{\bf 11}_2) \rightarrow (i=10{\bf 10}_2$, $j=11{\bf 11}_2)$.
    \item Convert $j$ into $j \ \text{XOR} \ i$ and then convert $i$ into $0000_2$.
    Do this by applying X gates to all that register qubits that encode the nonzero bits of $i$.
    Thus, ($i = {\bf1}0{\bf1}0_2$, $j={\bf1}1{\bf1}1_2) \rightarrow (i={\bf0}0{\bf0}0_2$, $j={\bf0}1{\bf0}1_2)$.
    \item Convert $j$ into $0001_2$ without changing $i$.
    For this purpose, we place CNOTs between the zeroth qubit (control) and the remaining qubits (target) that encode the nonzero bits of $j$.
    These CNOTs do not affect $i$ because in that part of the superposition, the control bit is zero.
    The final procedure yields $(i = 0{\bf0}00_2$, $j = 0{\bf1}01_2) \rightarrow (i = 0{\bf0}00_2$, $j=0{\bf0}01_2)$.
\end{enumerate}

\begin{figure}[ht]
    \noindent\centering{
    \includegraphics[width=0.9\columnwidth]{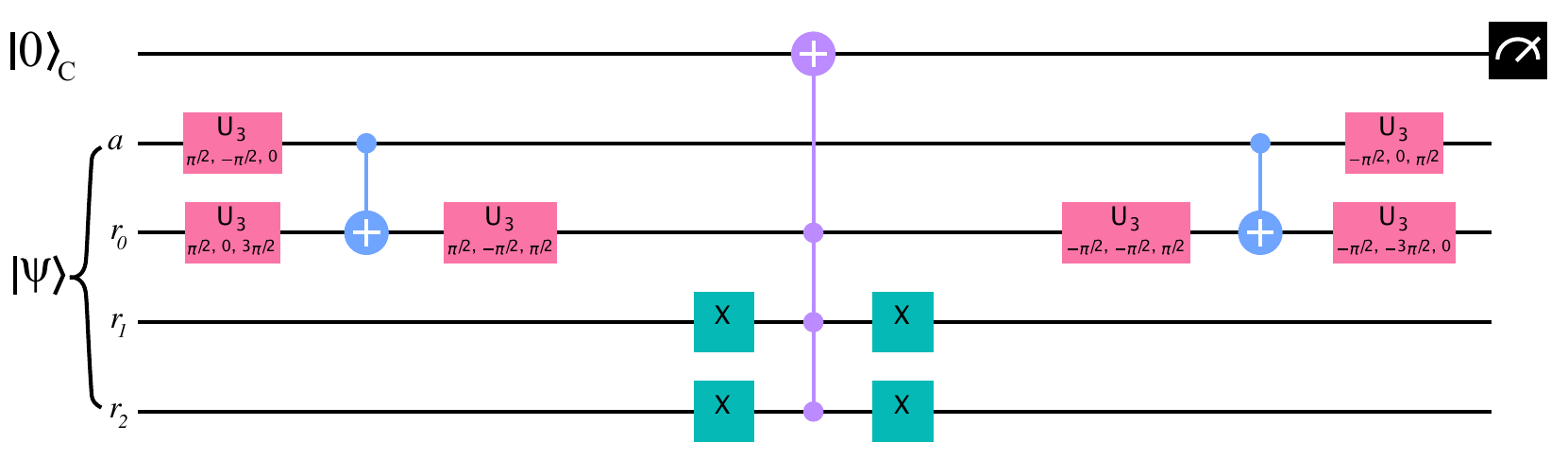} 
    }
    \caption{The circuit for separating states from $V$ and $V^*$ in the case where there are multiple constraints.
    }
    \label{fig:constraint_big_problem}
\end{figure}

After performing this algorithm, the circuit shown in Fig.~\ref{fig:meas_full_gate} can be applied.
However, 
it would spoil the optimization procedure because it affects not only $x_0$ and $x_1$ but also all the other $x_i$.
To avoid this, we replace the CNOT gate (that was between the constraint ancilla and the register qubit) with some X gates and a multi-qubit Toffoli gate as shown in Fig.~\ref{fig:constraint_big_problem}.
This results in an X operation being applied if and only if all the register qubits, apart from the zeroth one, are in the $\ket{0}$ state and the zeroth register qubit is in the $\ket{1}$ state.
This allows us to uncouple $x_0$ and $x_1$ from all other 
terms and implement the $x_0+x_1=1$ constraint.

%

\subsection{Multiple constraints}\label{subsec:MultipleConstraints}

Multiple $x_i+x_j=1$ constraints can be added sequentially to the circuit, matching each constraint with a new ancilla.
Accordingly, if we have an optimization problem with $N$ bits and $m$ constraints, the total number of qubits is $\lceil \log_2{N} \rceil + 1 + m$, where $\lceil...\rceil$ is the ceiling function.
For each constraint, we sequentially perform the circuit depicted in Fig.~\ref{fig:constraint_big_problem}. 

We measure all the ancillas, and when the outcome is 0, the state vector collapses into $P_nP_{n-1}\dots P_1\ket{\psi}$, where $P_i$ is a projection on feasible subspace $V_i$, corresponding to the $i^{th}$ constraint. It is known from linear algebra that $P_2P_1$ is a projection on $V_1 \cap V_2$ if and only if $P_2P_1=P_1P_2$. 

In the case where the constraints have no common bits, every pair of projections commute 
(see Appendix~\ref{AppendixSecCommuting})
and the state is projected onto the feasible subspace $V_1\cap V_2 \cap \dots \cap V_n$.

Using this scheme to encode constraints, it may be possible to encode more complex conditions, e.g., constraints with shared bits.
Also, linear constraints with multiple variables can be included in such a scheme.
These topics fall into the scope of future research.

\section{Methods}\label{sec:Methods}

In this section, we provide details about how we 
\textcolor{black}{simulate the QuEnc algorithm using classical hardware and how we run it on a quantum processor (see Sec.~\ref{subsec:ClassicalSimulation} and Sec.~\ref{subsec:QuantumImplementation} respectively).}

\subsection{Classical Simulation}\label{subsec:ClassicalSimulation}

The classical device used for the simulation is QMware HQC4020, which has 12 TB of RAM and 384 vCPUs in total.
All details on that hardware and its comparison in terms of the runtime and accuracy of specialized simulated and physical quantum processing units can be found in Ref.~\cite{kordzanganeh2022benchmarking}.
The amount of random access memory (RAM) utilized in simulating any noiseless quantum circuit grows exponentially with the number of qubits: an $n-$qubit state requires approximately $16 \times 2^n$ bytes of memory. 
For the 8,192-node graph (the largest problem that we solve here), we need a 14-qubit circuit, which equates to just 256 KB of RAM for a single circuit.
Thus, we can easily parallelize the gradient calculations and restarts of QuEnc.
  
\subsection{Quantum Implementation}\label{subsec:QuantumImplementation}

In order to probe the QuEnc algorithm on real quantum devices, we consider three sizes of {\tt MaxCut} problems: 4, 8, and 16 vertices. 
Efficient encoding allows us to use 3, 4, and 5 qubits, respectively, therefore any small-scale public quantum hardware can be used.

We utilize \textit{ibmq-manila} superconducting processor with the single-qubit error of $2.6\cdot10^{-4}$, an average CNOT error of $7.3\cdot10^{-3}$ and an average readout error of $2.5\cdot10^{-2}$.
For calculating gradients, we limit the number of measurements of a quantum state to 20,000.
In order to monitor the performance of the algorithm, we measure the normalized cost function, \(C_{norm}\), at each iteration of the gradient descent according to
\begin{equation}
    C_{norm}=\frac{C-C_{glob}}{C_{rand}-C_{glob}},
    \label{cnorm}
\end{equation}
where \(C_{glob}\) is the cost of global minimum and \(C_{rand}\) is the average cost of random solution. 
If $C_{norm} > 1$, the solution is worse than the random one, and if $C_{norm} = 0$ this implies that the global optimum is found.

\section{Results}\label{sec:Results}

In this section, we present the results of the unconstrained and constrained {\tt MaxCut} optimization problems \textcolor{black}{(see Sec.~\ref{subsec:UnconstrainedMaxCut} and Sec.~\ref{subsec:ConstrainedMaxCut} respectively)}. 
Unconstrained solutions are obtained on the \textit{ibmq-manila} superconducting device. However, in order for the performance of QuEnc to be compared to and hybridized with the IBM CPLEX solver, QuEnc is run on the classical QMware HQC4020 computer.

\subsection{Unconstrained {\tt MaxCut}}\label{subsec:UnconstrainedMaxCut}

\subsubsection{QuEnc simulation}\label{subsubsec:QuEncSimulation}

The performance of optimization algorithms often heavily depends on the graph structure.
As an illustrative example, we consider a simple graph where the first node is connected to all nodes, while there are no connections between other nodes.
The two solutions of this toy problem are $[1, 0, \cdots, 0]$ and $[0, 1, \cdots, 1]$, because the maximum cut isolates the first node.
While we know the optimal answer, the algorithm does not know these solutions and has to find one of these cuts.
The energy defined in Eq.~\eqref{energy} serves as a performance metric: lower energy indicates a better solution.
Considering random 8,192-node graphs, we easily find the exact solution to such problems using 14 qubits ($\log_2{8,192}=13$ register qubits + $1$ ancilla) with 4 layers, optimizing over 56 parameters via gradient-descent with ADAM optimizer.

\begin{figure}[ht]
    \noindent\centering{
    \includegraphics[width=70mm]{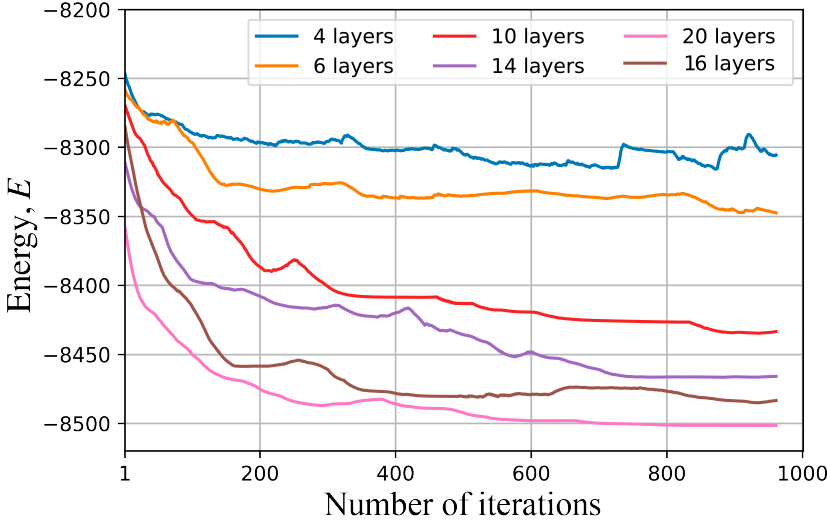}}
    \includegraphics[width=70mm]{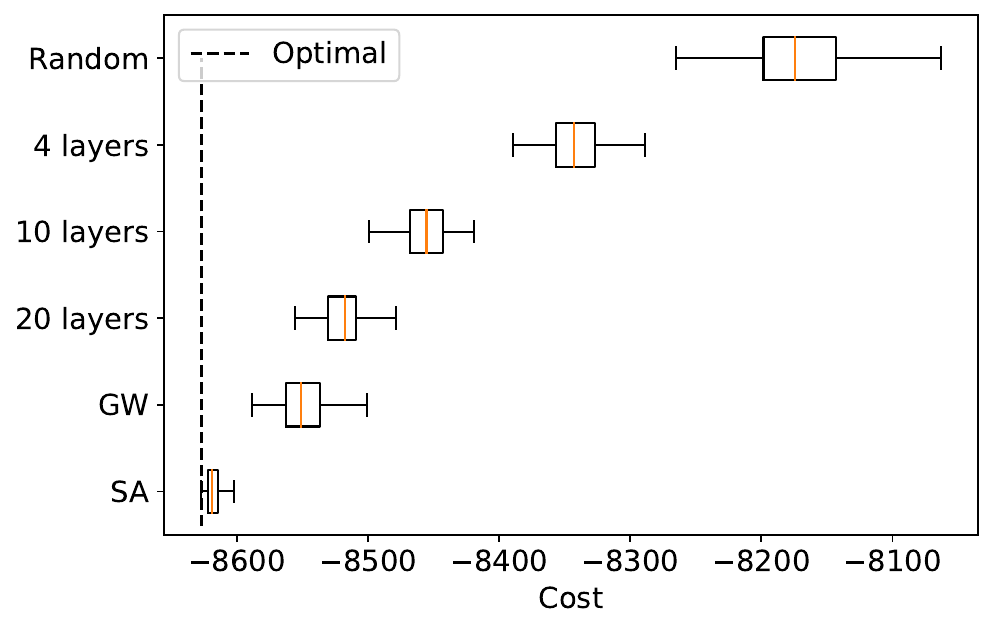}
    \caption{
    QCN performance with different numbers of layers on complete graphs with 256 nodes. (\textcolor{black}{left}) Convergence plot. (\textcolor{black}{right}) \textcolor{black} {Box plot of the cost after QuEnc convergence in comparison to the cost of a random bitstring, the Goemans-Williamson algorithm (GW), and simulated annealing (SA). 
    The optimal cost value is marked by the dashed line.}
    } 
    \label{complete_graph_256}
\end{figure}

In order to investigate the performance of our algorithm on more complex complete graphs where the solution is not known a priori, we define a randomly weighted, complete graph with 256 nodes.
The weights of the edges vary uniformly in the range $[0.01, 1]$.
The convergence of the energy of a 256-bit solution as a function of gradient descent iterations is shown in Fig.~\ref{complete_graph_256} (\textcolor{black}{left}), which illustrates the learning process of QuEnc.
Here, we tune the number of layers, change the circuit complexity, and show the performance of shallow circuits with between 4 and 20 layers.

\textcolor{black}{
In order to evaluate the performance of QuEnc as a heuristic optimization algorithm, it is compared with other heuristic approches, such as the Goemans-Williamson (GW) algorithm and simulated annealing (SA).
Besides, we use the IBM ILOG CPLEX software package~\cite{cplex2009v12} as an optimization approach that can prove optimality.
The box plot of the found solution costs for QuEnc with different number of layers, for GW, and for SA is shown in Fig.~\ref{complete_graph_256} (right).
The cost of the optimal solutions obtained with CPLEX and the cost of a random solution, where the cost function is computed for random bitstrings, is also plotted. 
The performance of QuEnc improves as the circuit depth increases, both in terms of the mean value and the uncertainty of the solution. 
The performance of QuEnc with 20 layers is close to that of GW.
Simulated annealing is able to find solutions close to the optimal ones.}

\subsubsection{Hybridizing QuEnc and CPLEX}\label{subsubsec:HybridizingQuEncAndCPLEX}

The IBM CPLEX solver, which has been developed for more than three decades~\cite{cplex2009v12}, is highly optimized and can be well-tuned to existing computing infrastructures.
However, the fine-tuning of such solvers to achieve more optimal solutions in a fixed runtime is a challenging process and requires certain internal optimization.
Here, we show that quantum algorithms can serve as a booster for classical solvers and help them to find the solution in less time.
Mainly, we resolve the problem using QuEnc, which obtains the solution at a decent cost and use that solution as the initial point for the CPLEX solver.
In this way, the solution from one algorithm/solver warm-starts the other. 
\begin{figure}[ht]
    \noindent\centering{
    \includegraphics[width=70mm]{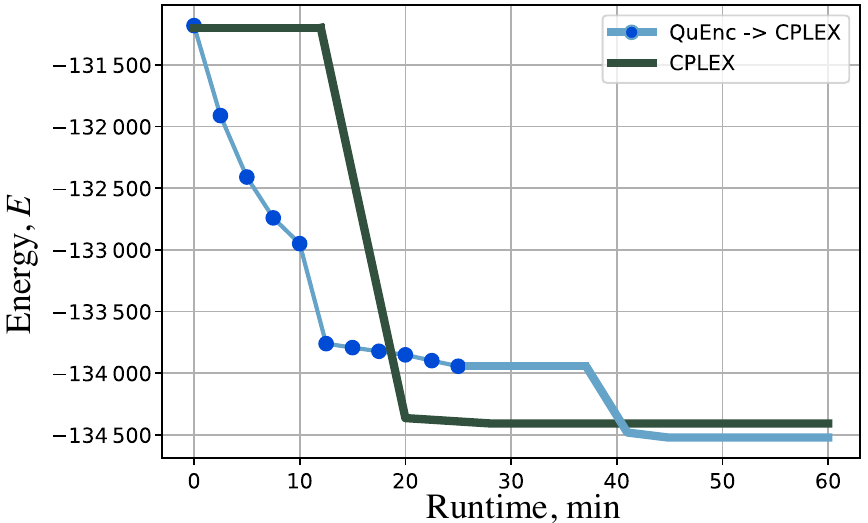}}
    \caption{
    The CPLEX and QuEnc$\rightarrow$CPLEX performance on a complete graph with 1024 nodes.
    } 
    \label{complete_graph_1024}
\end{figure}

In order to demonstrate the capabilities of such a hybrid pipeline, we consider a {\tt MaxCut} problem on a 1024-node graph, where the optimization landscape is very complex, generated in the same manner as before for the 256-node graph.
The QuEnc and CPLEX convergence (i.e., the reduction of the energy (cost) as a function of execution time) is presented in Fig.~\ref{complete_graph_1024}.
The blue dots show that the QuEnc solution converges at $-13,400$.
The QuEnc simulation with 11 qubits and 30 layers results in each iteration being around 3 minutes.
The light blue line is the convergence of the CPLEX, which starts from the point that was found by QuEnc. 
The plateau arises due to the CPLEX internal processes, such as branching tree building~\cite{cplex2009v12}.
The green line corresponds to the pure CPLEX solution started at a random point, whose cost is close to the initial cost of QuEnc, which also starts from a random point. 
Limiting the runtime to one hour, we find that the QuEnc$\rightarrow$CPLEX pipeline finds a better cost ($-134,520$) than CPLEX ($-134,406$), providing an improvement of 8.5 basis points. 
All experiments were conducted using QMware HQC4020~\cite{kordzanganeh2022benchmarking}.
{\color{black} Here, the obtained improvement may seem quite small. However, for large-scale optimization problems that are solved on a regular basis and must be processed within a fixed time, for instance multi-billion portfolio optimization, such an improvement could play a crucial role and lead to huge cost savings.

We would like to emphasize that such an improvement depends on the analyzed problem and the data as well as on the classical optimization technique \cite{Somov2023}.}
Theoretically, more careful tuning of the classical solver may provide better solutions, but practically, such tuning requires significant effort and resources.
Moreover, enhancing the quantum solution, especially implemented using dedicated hardware, allows for improving the time-to-solution for discrete optimization problems.
Experiments on a specific {\tt MaxCut} problem, demonstrated here, serve as an indication of the potential of the hybrid pipelines and pave the way toward future hybrid developments.

\subsubsection{QuEnc run via NISQ device}\label{subsubsec:QuEncRunViaNISQdevice}

\begin{figure}[ht]
	\noindent\centering
	\includegraphics[width=80mm]{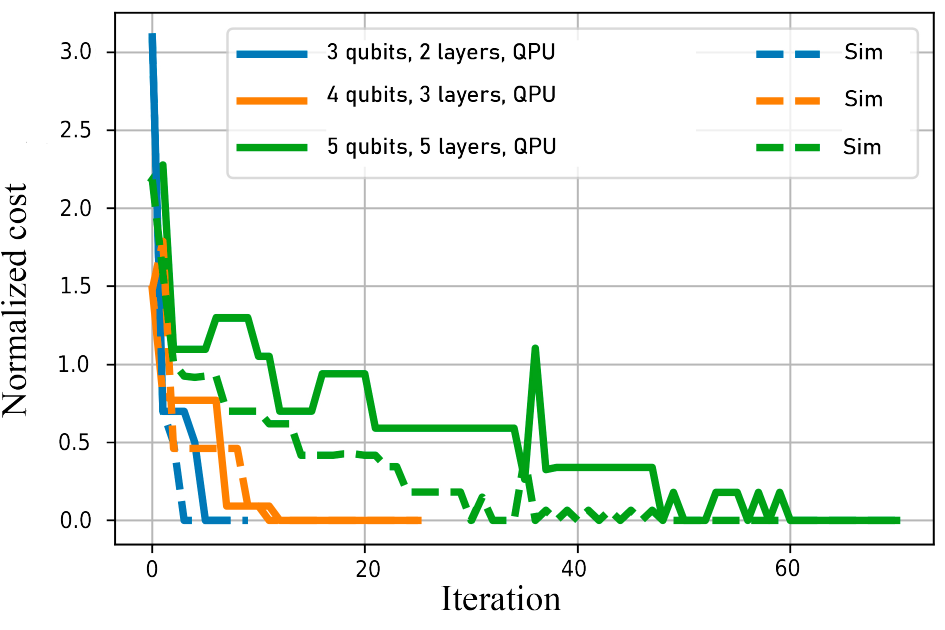}
	\caption{Learning process {\color{black} (convergence plot of the cost defined in Eq.\,\eqref{cnorm})} on the QPU (solid lines) in comparison with the numerical simulation (dashed lines) from the same initial point. 
    {\color{black} The normalized cost is depicted as it allows for the visual comparison between different sizes of the problems and demonstrates that the QPU is able to solve all problems to the global optima.}
		Each color corresponds to a certain problem size, which is indicated on the legend: 3 qubits for 4 vertices, 4 qubits for 8 vertices, and 5 qubits for 16 vertices. 
		QAOA requires 4, 8, and 16 qubits, correspondingly, to solve the same problems.
	}
	\label{fig:qpu_results}
\end{figure}

The convergence of the QuEnc run on the superconducting device is shown in Fig.~\ref{fig:qpu_results} for three {\tt MaxCut} problems.
Different colors correspond to different sizes of the considered graph.
While solid lines show the experimental results, dashed lines are obtained by simulating the quantum circuit with the same initial settings.
Due to the hardware-efficient structure and shallowness of the circuit, the NISQ device is able to find global optima and demonstrate similar convergence as the simulated algorithm.

These results indicate the possibility to utilise NISQ devices with limited topology and undesired errors for solving optimization problems.
Larger-scale implementation, which is required to study the algorithm further, is a subject of future research.

\subsection{Constrained {\tt MaxCut}}\label{subsec:ConstrainedMaxCut}

We solve the constrained {\tt MaxCut} problem using QuEnc without any classical pre- or post-processing.
Constraints are realized directly in a quantum circuit of QuEnc, and the convergence to solutions that violate posed constraints is forbidden by the quantum state engineering as described above.
\begin{figure}[ht]
    \centering
    \includegraphics[width=150mm]{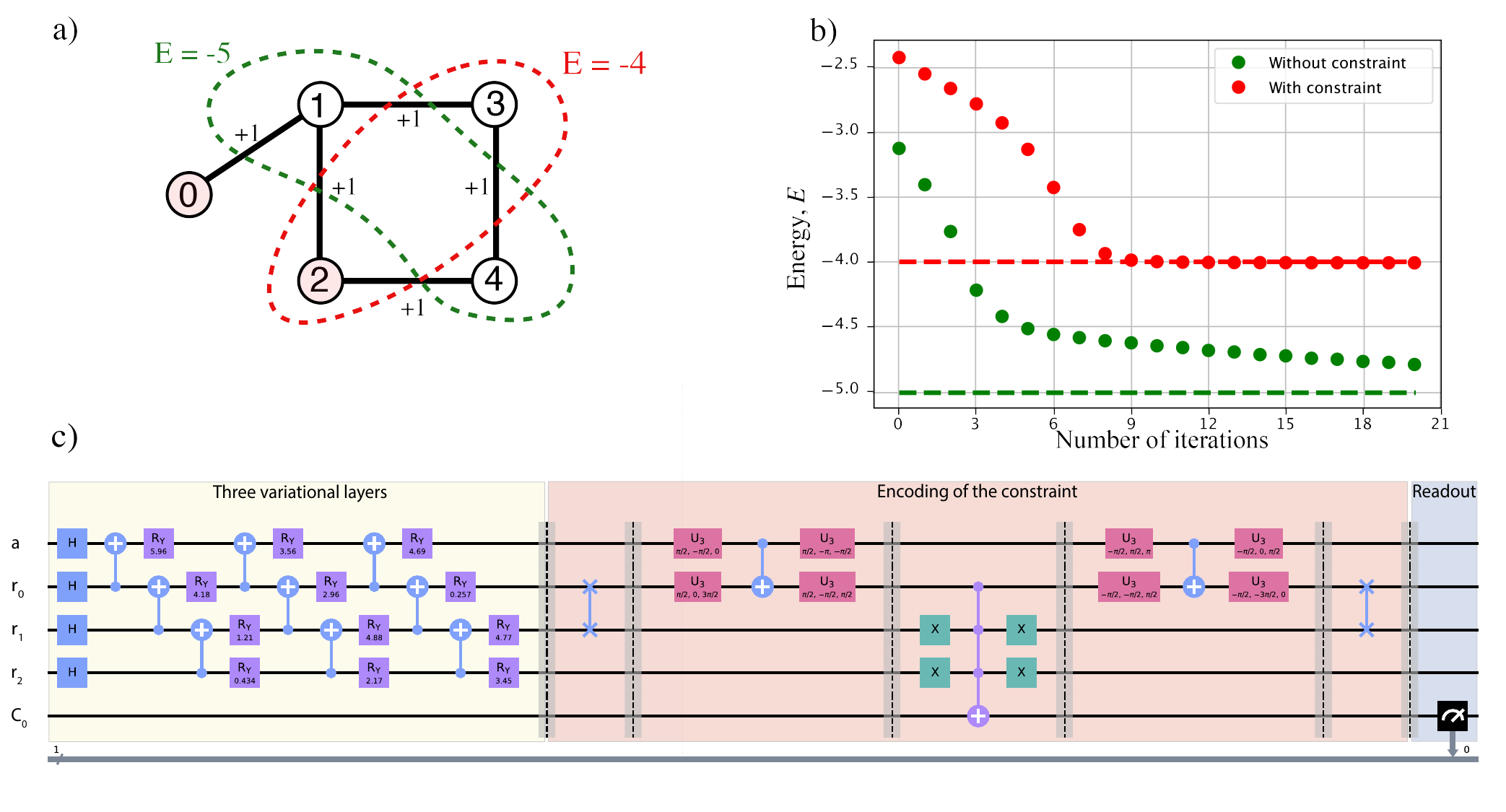}
    \caption{
    (a) The unweighted {\tt MaxCut} problem graph, where for the constrained problem, the two red vertices (i.e., the $0^{th}$ and the $2^{nd}$) must be in different subgroups.
    The number of edges that connect the vertices of different subgroups is the value of the cut, which is to be maximized.
    When there are no constraints, the maximal cut (indicated by the green dashed line) has a value of -5 since it involves all the edges.
    The maximal cut that satisfies the specified constraint is shown by the red dashed line and has a value of -4.
    (b) A plot of the energy (or cost) of the solutions as a function of the number of iterations. 
    The constrained solutions are shown in red and the unconstrained solutions are in green.
    The dashed lines show the cost of the optimal solutions.
    Even though the green dots do not reach the dashed line, the most probable solution after 6 iterations is the optimal one.
    {\color{black} (c) Full quantum circuit including variational layers, which were optimized for the given graph (in yellow rectangular), and circuit of the constraint encoding (in red rectangular). 
    Given quantum circuit provides the solution for the 5-node constrained {\tt MaxCut} problem.}
    }
    \label{fig:constraint_experiment}
\end{figure}

In order to illustrate the effect of the constraints, we consider an unweighted graph  ($d_{ij}$ is either 1 or 0) with 5 nodes, depicted in Fig.~\ref{fig:constraint_experiment}(a).
Such a graph has the obvious maximum cut shown using a green dashed curve.
The energy value is $-5$ since the cut goes through all the edges and the minus sign appears due to the cost formulation in Eq.~\eqref{energy}.
The 4-qubit QuEnc finds the solution of this problem using 5 layers as shown in Fig.~\ref{fig:constraint_experiment}(b): green dots represent the algorithm's convergence and the most probable solution at the last iteration yields exactly $-5$.

While in the previous examples there were no constraints on a cut, now we add an additional condition such that the $0^{th}$ and $2^{nd}$ vertex must be in different subgroups{\color{black}, i.e., $x_0+x_2=1$.}
This constraint does not allow achieving the maximum cut, which includes 5 edges and contains no more than 4 edges (the cost is $-4$).
The optimal cut, in that case, is shown in Fig.~\ref{fig:constraint_experiment}(a) using a dashed red curve.

We probe the QuEnc algorithm with the same number of layers but with a constraint, which adds an additional qubit to the circuit.
It finds a solution (with vertices 0, 1, and 4 allocated to the first subgroup and vertices 2 and 3 allocated to the second subgroup) that satisfies the constraint and at the same time is the maximum cut.
The QuEnc convergence is shown in Fig.~\ref{fig:constraint_experiment}(b) in red.
We note that there is better convergence in this constrained case than in the unconstrained one.
{\color{black} The optimized quantum circuit with constraint encoding part is shown in Fig.~\ref{fig:constraint_experiment}(c)}.

\section{Analysis}\label{sec:Analysis}

In order to gain insight into our quantum algorithm, we analyze it in several ways.
First, we investigate the problem of convergence to sub-optimal solutions due to the presence of local minima \textcolor{black}{(see Sec.~\ref{subsec:LocalMinima})}. 
We proceed to analyze how the output of the QuEnc algorithm depends on the number of shots required by the probabilistic nature of quantum measurements \textcolor{black}{(see Sec.~\ref{subsec:QuantumStateMeasurement})}.
We consider the possibility of efficiently simulating QuEnc with tensor networks \textcolor{black}{(see Sec.~\ref{subsec:SimulationAnalysis})} and also provide a scaling analysis of the algorithm \textcolor{black}{(see Sec.~\ref{subsec:Scaling})}.
Finally, we examine whether an alternative ansatz for the quantum circuit of the QuEnc algorithm achieves better performance on {\tt MaxCut} problems and compare their expressibility \textcolor{black}{(see Sec.~\ref{subsec:AnsatzAnalysis})}. 

\subsection{Local minima}\label{subsec:LocalMinima}

Local minima prevent QuEnc from finding the global optimum, which results in a sub-optimal solution.
The easiest way to investigate local minima is to restart the algorithm with a random initialization multiple times.
Here, analysis is carried out for the {\tt MaxCut} problem on a noiseless simulator. 
We launch QuEnc for $1000$ random problems of the same size from random initial points until convergence and analyze the probability of finding the global optimum. 
The experiment is repeated for different numbers of layers and problem sizes with the number of gradient decent steps required for convergence.
\begin{figure}[ht]
     \noindent\centering
    \includegraphics[width=80mm]{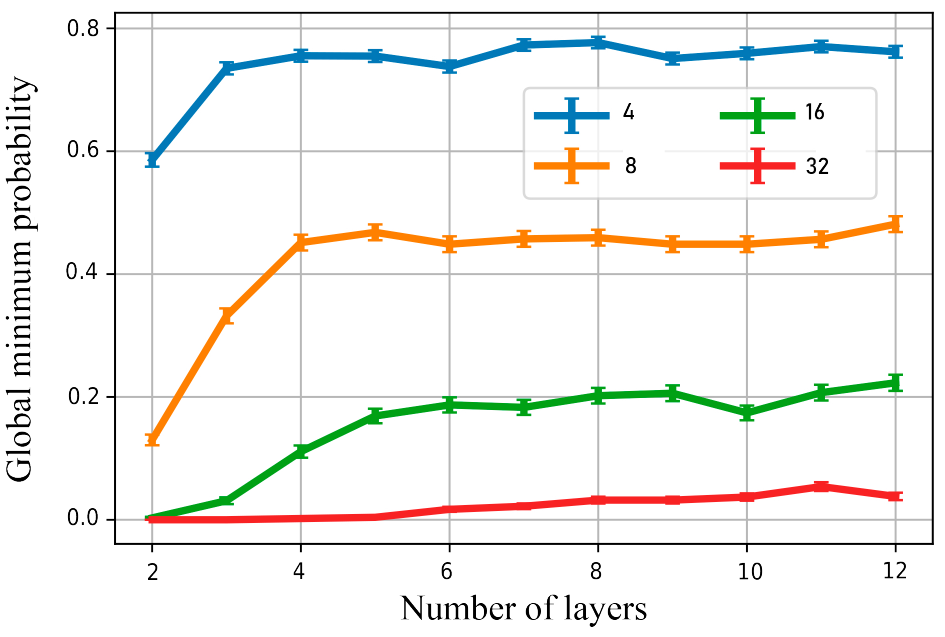}
    \caption{Probability of finding global minimum for unconstrained {\tt MaxCut}.
    Gradient descent.
    Noiseless simulation.
    Each curve is obtained by averaging over 1000 launches and corresponds to a certain problem size, which is indicated on the legend. 
    }
    \label{fig:local_minima}
\end{figure}

The results of this analysis are presented in Fig.~\ref{fig:local_minima}.
We investigate random graphs with 4, 8, 16, and 32 nodes and plot the average probability to find the global minimum as a function of circuit depth.
It is clear that a higher number of layers helps the global optimum to be found more frequently, but the probability saturates.
For a 32-node {\tt MaxCut}, QuEnc with 6 qubits and 11 layers finds the correct cut with 7\% probability.
Larger problem analysis requires more powerful hardware since the circuit evaluation has to be repeated multiple times and averaged over random graphs.

The variational quantum algorithm has to be restarted multiple times in order to avoid local minima.
The efficient parallelization of this algorithm is a key component in a search for global optima, which falls into the scope of future research. 

\subsection{Quantum state measurement}\label{subsec:QuantumStateMeasurement}
\newcommand{\C}{\left<C\right>}

The nature of quantum computation does not allow for an immediate full quantum state readout.
Even with a fault-tolerant device, we would be limited in practice by the number of measurements (since the overall runtime required to obtain the solution is proportional to the number of these shots).
In order to estimate the effect of the number of shots, we launch the QuEnc algorithm for different random {\tt MaxCut} problems of the same size until convergence and then calculate the average cost, $\C$.
We analyze this cost as a function of the problem size, $n_c$, the required learning rate, $\alpha$, and the number of shots, $k$, denoting it as $\C(n_c, \alpha, k)$.
In the case of $k\to \infty$, we use the full-state vector simulation. 

The learning rate $\alpha_0$ is set so as to minimize the averaged over different graphs cost function $\C(n_c, \alpha_0, \infty)$.  
This value is unreachable in a real experiment, and we use it as a reference level.
We would like to emphasize that the infinite number of shots does not provide the optimal solution -- we do not pick the best solution out of experiments but rather average over all the experimental results.

Considering a realistic scenario with a limited number of shots $k$, we adjust the learning rate $\alpha$ to mitigate errors caused by shots and 
calculate $\C(n_c, \alpha, k)$.
In order to compare the introduced cost functions, we analyze the relative cost
    \begin{equation}
        \widetilde{\C}(n_c, \alpha, k) = \frac{\C(n_c, \alpha, k)-\C(n_c, \alpha_0, \infty)}{\left<C_{rand}\right>(n_c) - \C(n_c, \alpha_0, \infty)},
        \label{eq:rel_cost}
    \end{equation}
where $\left<C_{rand}\right>$ is the average cost of a random bitstring.
When the solution is no better than a random bitstring, $\widetilde{\C}(n_c, \alpha, k)$ is 1, and when the cost functions for the finite and infinite number of shots coincide, it is 0.

\begin{figure}[ht]
    \noindent\centering
    \includegraphics[width=80mm]{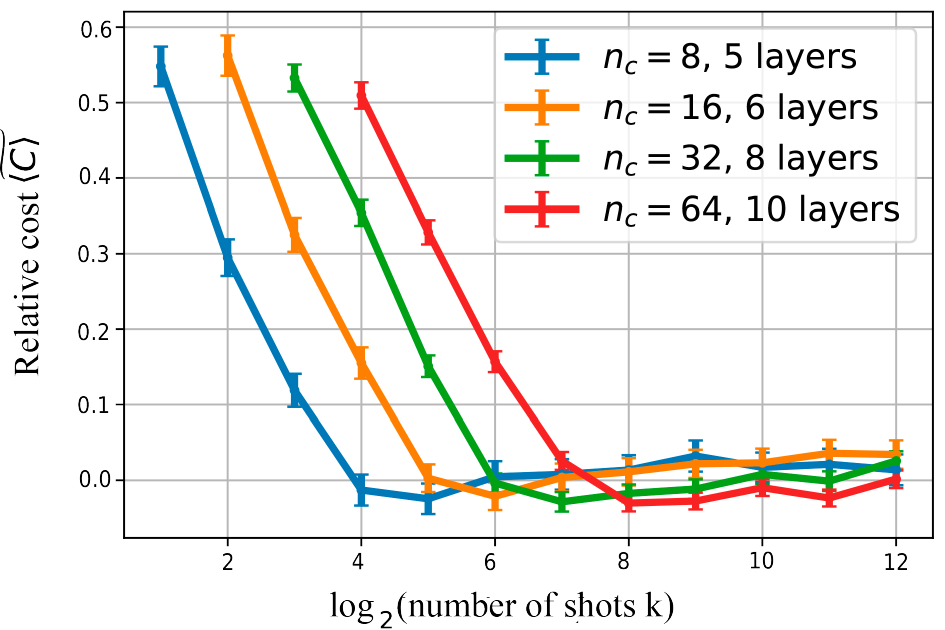}
    \caption{QuEnc efficiency with different numbers of shots for several problem sizes. 
    Relative cost, defined in Eq.~\eqref{eq:rel_cost}, as a function of the number of shots in log-scale. 
    The zero level corresponds to learning with $k=\infty$. 
    All curves were obtained by averaging over $\sim 250$ problems.}
    \label{fig:shots2}
\end{figure}

The relative cost as a function of the number of shots averaged over 250 fully-connected, weighted graphs is presented in Fig.~\ref{fig:shots2} with different graph sizes indicated in different colors.
We adjust the number of layers for each problem size and the learning rate for each value of $k$ 
(see Appendix~\ref{AppendixSecLearning}).
Remarkably, some finite values of $k$ provide a better averaged cost than $k\to \infty$, as indicated by negative values of the relative cost.
By limiting the number of measurements, we introduce the additional stochasticity to our gradient descent that helps to avoid local minima~\cite{Liu2022Noise}.
From the presented experiments, we find that the required number of measurements scales linearly with the small problem size: $\widetilde{\C}<0$ in the case of $k\approx4\,n_c$. 
Scaling for larger problems will be evaluated in the future.

\subsection{Simulation analysis}\label{subsec:SimulationAnalysis}

The hardware-efficient quantum circuit, where only neighboring qubits are connected, allows us to utilize not only the features of the NISQ devices but also the efficient simulation algorithms based on tensor networks. 
Tensor networks represent a mathematical tool initially created for solving multi-particle problems in quantum physics~\cite{DMRG} but
are now actively used for quantum circuit simulations~\cite{Cotengra, Markov_simulation}.
For instance, Ref.~\cite{close_supremacyAndOthers} presents results on the simulation of Google's well-known quantum supremacy circuit~\cite{AruteAndOthers2019} with fixed accuracy at a very high speed. 
Moreover, as was shown in Ref.~\cite{MPS_preparation2}, quantum circuits with structures similar to QuEnc's generate states corresponding to the Matrix Product State (MPS) class, which is well studied.

Tensor network simulation allows for the analysis of the circuit expressibility and entanglement with respect to the number of layers~\cite{Nakaji2021}. 
It is important to monitor how the MPS rank $r$ (bond dimension)~\cite{oseledets2011tensor_train} of the output state of the quantum circuit increases with the number of layers. 
The rank determines the entanglement of the state, since 
the entropy of the reduced state obeys the bound $S \leq \log r$ -- details can be found in Ref.~\cite{orus2014practical}.
Moreover, utilizing the tensor networks-based simulation, we can realize the pre-learning of shallow circuits and then insert the resulting state in a quantum computer as an initial point, which allows us to implement a warm-start optimization (see Appendix~\ref{AppendixSecWarm}).

%
\begin{figure}[ht]
    \centering
    \includegraphics[width=70mm]{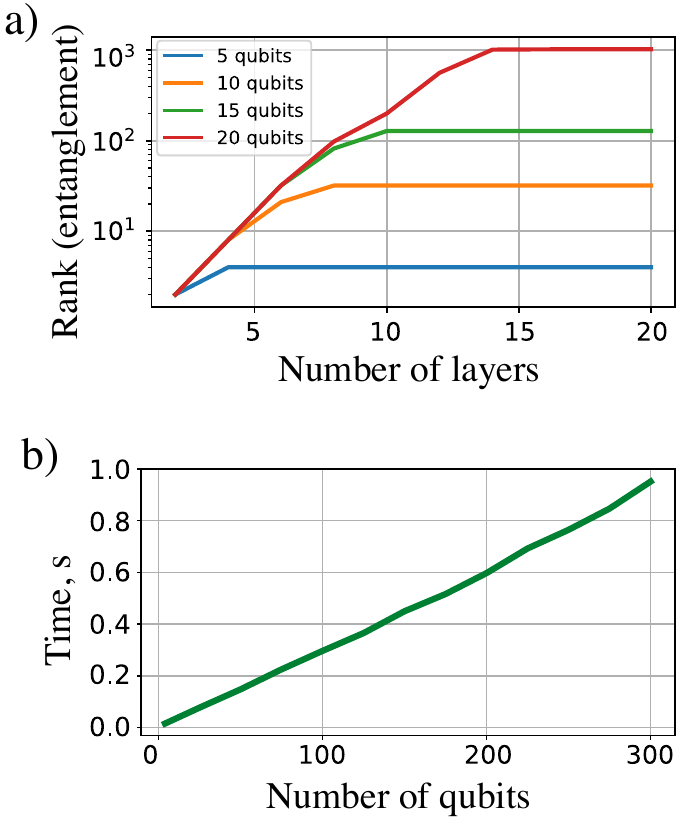}
    \caption{(a) The \textcolor{black}{maximum} MPS rank \textcolor{black}{at half system size bipartition}, which characterizes the entanglement of the output state of the QuEnc circuit as a function of the number of layers, $L$, for different numbers of qubits, $n$.
    For each value of $n$, the depth for which the output state has a full MPS rank \textcolor{black}{(equal to $2^{n/2}$)}, i.e. maximum entanglement, scales as $L \sim \frac{3}{4} n$.
    (b) Simulation time of a single circuit with 5 layers projected onto the state $\ket{0}^{\otimes n}$ as a function of $n$. Low entanglement allows us to simulate a 300-qubit circuit in less than a second.}
    \label{fig:TN_qubits}
\end{figure}

We analyze the rank of the corresponding circuit for the different numbers of qubits as a function of the number of layers in Fig.~\ref{fig:TN_qubits}(a).
A high rank does not allow us to simulate circuits efficiently using tensor networks, however, shallow circuits provide a low rank and can be processed.
We simulate a shallow 5-layer QuEnc circuit with up to 300 qubits, demonstrating the linear scaling with the number of qubits in case of a fixed number of layers, see Fig.~\ref{fig:TN_qubits}(b). 
It is worth noting that we measure the time it takes to obtain either the output state in the MPS format or the projection onto any basis vector $\ket{i}$ and not the full state vector.
A detailed description of finding tensor network contraction sequences is presented in Appendix~\ref{AppendixSecContraction}.

In comparison with the large-scale QAOA tensor simulation~\cite{Lykov2020TensorNQ}, we are able to investigate wider circuits without a high-performance supercomputer by using just a regular laptop.
This is due to the simple structure of the QuEnc circuit and the low levels of entanglement that are sufficient to provide a solution.
The limitation of such an approach is the number of layers since tensor networks can efficiently simulate only shallow circuits, which might not be enough for large-scale problems; otherwise, the ranks are too high.

\textcolor{black}{Regarding the actual launch of this algorithm on a quantum computer, it is worth emphasizing that it makes sense to do this only with a large number of qubits ($n \gtrsim 24$) and a large number of layers ($ \gtrsim n $) because in any other cases it is much more efficient to simply simulate it using tensor networks. 
At the same time, due to the presence of the barren plateau problem in hardware-efficient circuits, various techniques should be used to combat this, e.g., warm-start (see Appendix \ref{AppendixSecWarm}).}

\subsection{Scaling}\label{subsec:Scaling}
\textcolor{black}{Recall from Sec.~\ref{sec:Learning} that one step of gradient descent requires $2n_q\times L+1$ experiments, where $n_q=O(\log n_c)$ is the number of qubits and $L$ is the number of layers. 
Therefore, a single QuEnc launch requires $O(\log n_c\times L\times k \times T)$ experiments for $T$ iterations of gradient descent, where $n_c$ is the size of the QUBO.
Here, $k$ is the total number of shots for each circuit, which we assume scales linearly with the size of the problem.
With $L$ scaling as $O(\log{n_c})$, which is enough to achieve high entanglement and expressibility~\cite{luchnikov2021riemannian}, the runtime of one experiment for the Sequential-2QG circuit scales as $O(\log^2 n_c)$, because its depth scales not only with the number of layers but also with the number of qubits. 
As a result, we estimate the runtime of a single QuEnc launch as $O(T\times n_c\log^4 n_c)$.}
Here, we do not take into account the number of restarts of the algorithm since smart circuit initialization, e.g., warm-start (see Appendix~\ref{AppendixSecWarm}), may be used to solve large-scale problems and a separate scaling analysis is required.

{\color{black} Since a hardware-efficient circuit is employed, the barren plateau problem exists in our setup \cite{Larocca2022}.
This can be addressed by utilizing smart circuit initialization or cutting the quantum circuit in similar manner as in Ref.~\cite{Melnikov2023}.
Since a logarithmic number of qubits is used, large-scale optimization problems can be addressed with relatively low-width circuits.}
In a parallel quantum computing scenario, where each gradient can be evaluated in a single run and all restarts can be performed in parallel (for instance, using a large multi-qubit chip), the runtime may achieve \textcolor{black}{$O(T \times n_c \log^2 n_c)$} scaling.

\subsection{Ansatz analysis}\label{subsec:AnsatzAnalysis}

In general, the ansatz has to be defined by the problem structure.
Due to the limited connectivity of the hardware, we are restricted by the hardware-efficient circuits. 
The quality of the circuit is defined by the expressibility~\cite{Nakaji2021, yczkowski2005} and related learning capability~\cite{holmes2022connecting}.
In order to be able to represent all classical solutions, among which we perform the search, the circuit has to feature high expressibility. 
Expressibility shows that the level of completeness and uniformity of the output state space of the circuit is covered. 
See Appendix~\ref{AppendixSecExpressibility} 
for a formal definition of expressibility.

\begin{figure*}
	\centering
	\includegraphics[width=0.85\textwidth]{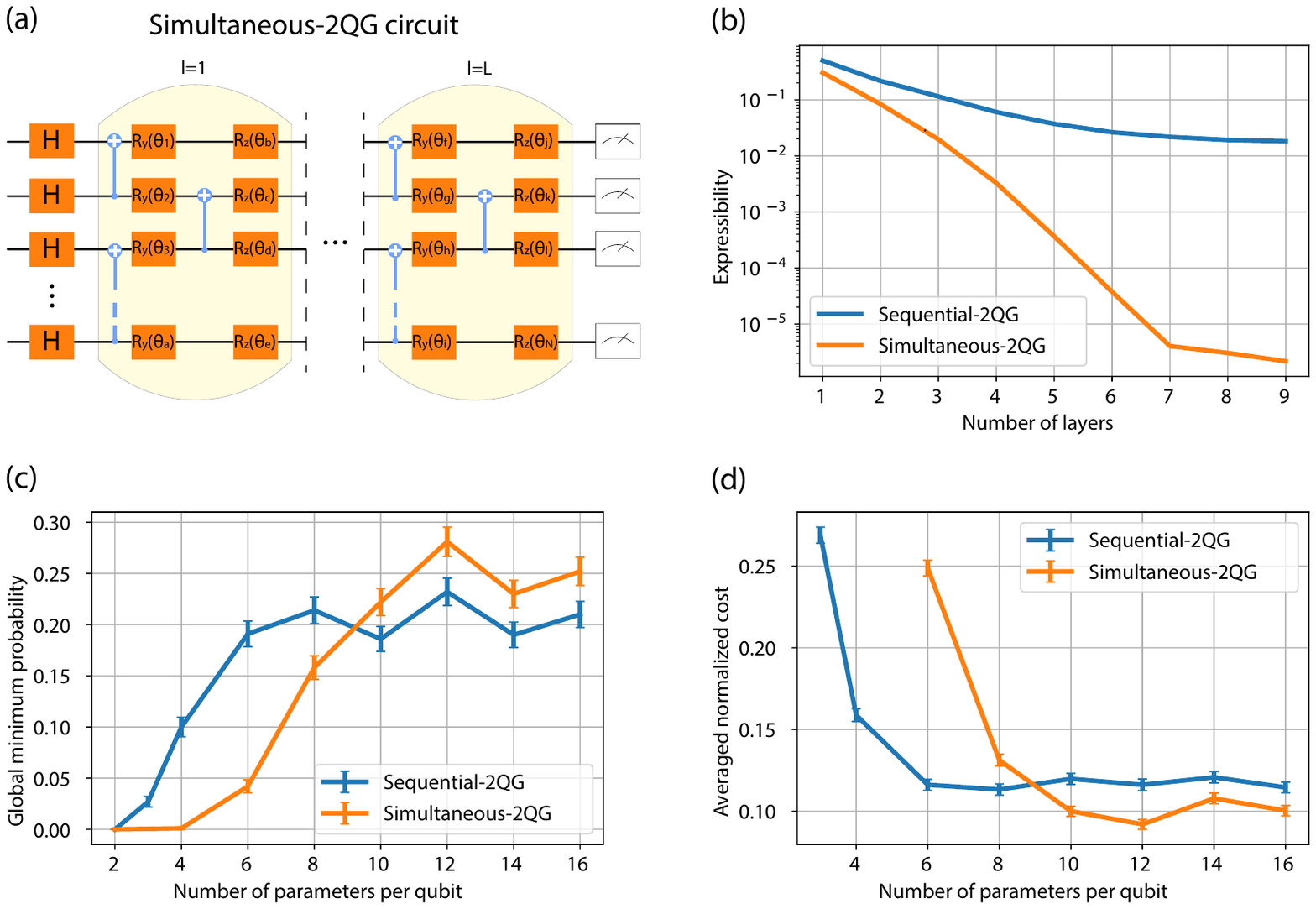} 
	\caption{
 (a) Simultaneous-2QG circuit -- new circuit, which is compared with our main circuit used in this paper (Sequential-2QG circuit).
    (b) Comparison of their expressibilities [Eq.~\ref{eq:q_expr}] on 9 qubits. 
    (c) Probabilities of finding the global minimum of 16-node {\tt MaxCut} problem that is solved using 5 qubits. 
    Data is obtained from 1000 launches at each point. 
    (d) Averaged normalized cost after learning in the same conditions as in (c). 
    } 
	\label{fig:step_circuit}
\end{figure*}

Here, we analyze two hardware-efficient quantum circuits, where the 2-qubit gates (2QG) are performed either sequentially (Sequential-2QG) or in parallel (Simultaneous-2QG).
The essential difference is that while the circuit depth of the former ansatz (shown in Fig.~\ref{qcn}) scales linearly with the number of qubits, the depth of the latter circuit (shown in Fig.~\ref{fig:step_circuit}(a)) is independent of the number of qubits. 
Note, however, that slightly different single-qubit gate sets are used in the two circuits.

The expressibility of the Simultaneous-2QG circuit significantly exceeds the expressibility of the Sequential-2QG circuit as can be seen in Fig.~\ref{fig:step_circuit}(b) where we show the example of a 9-qubit circuit.
Interestingly, there is a strong correlation between the expressibility of the quantum state and the classical representative power of the model (see Appendix~\ref{AppendixSecExpressibility} 
for further details).
We also calculate the Kullback–Leibler divergence between the bitstrings generated at the output of the circuit and the uniform distribution and found the same dynamics with respect to increases in circuit depth.

Despite the fact that the Simultaneous-2QG features much better expressibility, the probability of finding the global optimum and the average cost function in the context of {\tt MaxCut} problem does not improve so vastly, see Fig.~\ref{fig:step_circuit}(c, d) where we depict these values as functions of variational gates per single qubit (effective layers).
We suppose that this performance is related to the learning capability of the model.
Further investigation and ansatz analysis are required in order to adjust the quantum circuit to specific problems.

\section{Conclusion}\label{sec:Conclusion}

We propose a variational quantum algorithm for near-term gate-based devices that aims at the solution of discrete optimization problems. 
We propose a technique for directly encoding simple constraints using modifications of the quantum circuit instead of cost function crowding.
This technique is applicable to both probability encoding and one-hot encoding in the QAOA.

The advantage of the QAOA is the constant number of measurements that need to be completed in order to read out the binary string due to the one-hot encoding.
In contrast, the QuEnc encodes the problem in probabilities that lead to the linear number of measurements in the number of classical variables.
However, the QuEnc may be more suitable for the current NISQ devices where the number of high-quality qubits is restricted while the number of measurements can be arbitrary. 

We test the algorithm using a high-performance QMware universal and tensor networks-based simulator as well as a superconducting QPU.
The analysis of the algorithm scaling in the number of quantum state measurements and different initial points for optimization is presented.
Considering complete weighted {\tt MaxCut} problems,
we demonstrate that the QuEnc algorithm is a promising candidate for tackling binary optimization on a large scale (1000 nodes) whenever quadratic models appear via merging it with the classical algorithm in a hybrid solution.
After a thorough analysis, we conclude that the presented results serve as experimental evidence that the proposed approach may allow optimization applications on intermediate-scale quantum devices to be realized sooner than other variational algorithms.
Besides, we tested our algorithm in workflow scheduling problems within a hybrid pipeline in Ref.~\cite{pakhomchik2022solving}.\\

The authors have recently become aware of Ref.~\cite{novel_var_arg_published, novel_var_arg2} that also use exponentially fewer qubits compared to QAOA. 
However, those schemes are based on the Laplacian matrices for {\tt MaxCut}, specifically unconstrained optimization and classical optimization techniques, while in our work, we utilize a quantum method for the gradient evaluation for an arbitrary QUBO and use an ancillary qubit scheme that allows for capturing certain correlations.

\begin{acknowledgments}
We thank Dr. Margarita Veshchezerova, Dr. Alexey Melnikov, Vishal Shete, and Karan Pinto for their valuable discussions and suggestions.
\end{acknowledgments}

\newpage

\appendix

\section{Unconstrained Optimization}

\subsection{Proof of the equivalence of the general Ising model and {\tt MaxCut}}
\label{AppendixSecProof}

In this work, we focus on a {\tt MaxCut} problem due to its simplicity and equivalence to the general Ising model, which covers many valuable combinatorial optimization problems. 

We prove the equivalence of the Ising model defined in Eq.~\eqref{Ising} and {\tt MaxCut} defined in Eq.~\eqref{Hg}.
The equivalence means that for any ground state $\vec{s}$ of $H$, there are exactly two ground states of $H_g$ (let us call them $\vec{S^1}$ and $\vec{S^2}$) and that $H_g$ has no other ground states.
The states $\vec{S^1}$ and $\vec{S^2}$ are connected with spins $\vec{s}$ in the following way:
\begin{equation}
\begin{aligned}
    & S^1_i = s_i, S^2_i = -s_i\ (\forall i \in I), \\
    & S^1_a = 1, S^2_a = -1,
\end{aligned}
\end{equation}
 
Let us prove that these states are indeed the only ground states of $H_g$.
We consider two different spin configurations of $H_g$ where $s_a = 1$ and $s_a = -1$.
We start with spin value $s_a = 1$.
In this case, the Hamiltonian $H_g$ is exactly the same as the Hamiltonian $H$.
Therefore, if $s_a = 1$, then the minimal value of $H_g(s_a = 1)$ is equal to the minimal value of $H$ with the same spin configuration of rest spins.
Another case is with spin value $s_a = -1$.
The Hamiltonian $H_g$ can be transformed in the following way:
\begin{equation}
\begin{aligned}
    H_g(s_a = -1) = - \sum_{i \in I} h_i s_i + \sum_{i \neq j, i, j \in I} J_{ij} s_i s_j = \\
    = \sum_{i \in I} h_i (-s_i) + \sum_{i \neq j, i, j \in I} J_{ij} (-s_i) (-s_j) = H(-\vec{s}).
\end{aligned}
\end{equation}
This equation shows that the minimal value $H_g(s_a = -1)$ is the same as $H$, but with all spins being flipped.

Let us consider again the Hamiltonian $H_g$ but without the linear part:
\begin{equation}
    H_g = \sum_{i \neq j} J_{ij} s_i s_j.
\end{equation}
Finding a ground state of the Hamiltonian without the linear part is equivalent to the problem of solving a {\tt MaxCut} in weighted graphs.
For demonstrating it, separate the spin with negative and positive values into two sets $V^-$ and $V^+$ respectively.
$E(V^-)$ and $E(V^+)$ are the sets of edges between the spins in one set.
$\delta(V^+, V^-)$ are the edges between the different sets and $E$ is set of all edges.
Then, the Hamiltonian $H_g$ can be transformed as follows
\begin{equation}
\begin{aligned}
    H_g &= \sum_{i \neq j} J_{ij} s_i s_j = \\
    &= \sum_{i, j \in E(V^+)} J_{ij}
    - \sum_{i, j \in \delta(V^+, V^-)} J_{ij} + \sum_{i, j \in E(V^-)} J_{ij} \\
    &= \sum_{i, j \in E} J_{ij} - 2 \sum_{i, j \in \delta(V^+, V^-)} J_{ij}.
\end{aligned}
\end{equation}
The term $\sum_{i, j \in E} J_{ij}$ does not depend on the spin configuration and always has the same value whereas the last term account for only weights of edges between different values of spins.
It is equivalent to maximizing weights in the {\tt MaxCut} problem where $w_{ij} = J_{ij}$~\cite{MaxCutIsingGlasses}.

\subsection{Derivative calculation}
\label{AppendixSecDerivative}

In order to compute the derivative in Eq.~\eqref{grad_calc}, we explicitly write down the Eq.~\eqref{paramshift} for a given state:

\begin{multline*}
    \partial_{\theta_j} \bra{\psi(\vec{\theta})} \left(\ket{1}\bra{1}\otimes\ket{i}\bra{i}\right) \ket{\psi(\vec{\theta})} =\\
    \frac{1}{2} \Big(
    \bra{\psi(\theta_j + \pi/2)} \left(\ket{1}\bra{1}\otimes\ket{i}\bra{i}\right) \ket{\psi(\theta_j + \pi/2)}
    - \\
    \bra{\psi(\theta_j - \pi/2)} \left(\ket{1}\bra{1}\otimes\ket{i}\bra{i}\right) \ket{\psi(\theta_j - \pi/2)}
    \Big).
\end{multline*}
As mentioned in the main text, we use the parameter-shift rule by shifting the parameter of the wave functions in measurement expression by $\pm \frac{\pi}{2}$ as follows:

\begin{multline}
    \partial_{\theta_j} \bra{\psi(\vec{\theta})} \ket{i}\bra{i} \ket{\psi(\vec{\theta})} =\\
    \frac{1}{2} \Big(
    \bra{\psi(\theta_j + \pi/2)} \ket{i}\bra{i} \ket{\psi(\theta_j + \pi/2)}
    - \\
    \bra{\psi(\theta_j - \pi/2)} \ket{i}\bra{i} \ket{\psi(\theta_j - \pi/2)}
    \Big).
\end{multline}

By using the chain rule for function composition, we get
\begin{multline}
    \partial_{\theta_j} |b_i(\vec{\theta})|^2 = \frac{\partial_{\theta_j} \bra{\psi(\vec{\theta})} \left(\ket{1}\bra{1}\otimes\ket{i}\bra{i}\right) \ket{\psi(\vec{\theta})}  } {|\braket{\psi(\vec{\theta})| i}|^2}~ -\\ \frac{\bra{\psi(\vec{\theta})} \left(\ket{1}\bra{1}\otimes\ket{i}\bra{i}\right) \ket{\psi(\vec{\theta})}  }  {  |\braket{\psi(\vec{\theta})| i}|^4  }~ \partial_{\theta_j} \left( \braket{\psi(\vec{\theta})||i}\braket{i||\psi(\vec{\theta})} \right),
\end{multline}
which is used to compute the gradient of the objective function.

\subsection{ADAM optimizer}\label{AppendixSecADAM}

Instead of adapting the parameter learning rates based on the average first moment (the mean) as in the RMSProp, the ADAM optimizer also makes use of the average of the second moments of the gradients (the uncentered variance).
Overall, it acts upon the gradient component by using the exponential moving average of gradients $m$ that helps to overcome the noise, and the learning rate component by dividing the learning rate $\alpha$ by the exponential moving average of squared gradients $v$, which also optimizes the speed of learning.
The parameter update rule with the ADAM optimization is given by
\begin{equation}
\begin{aligned}
    \vec{\theta}_{t+1} &= \vec{\theta}_{t} - \frac{\alpha}{\sqrt{v_t}+\epsilon} m_t, \\
    m_t &= \frac{\beta_1\,m_{t-1} + (1-\beta_1)\nabla \mathcal{C}|_{\vec{\theta}=\vec{\theta}_{t}}}{1-\beta_1^t}, \\
    v_t &= \frac{\beta_2\,v_{t-1}+(1-\beta_2)\left(\nabla \mathcal{C}|_{\vec{\theta}=\vec{\theta}_{t}}\right)^2}{1-\beta_2^t}.
\end{aligned}
\end{equation}

\subsection{Learning rate tuning}
\label{AppendixSecLearning}

\begin{figure}[ht]
    \noindent\centering
    \includegraphics[width=50mm]{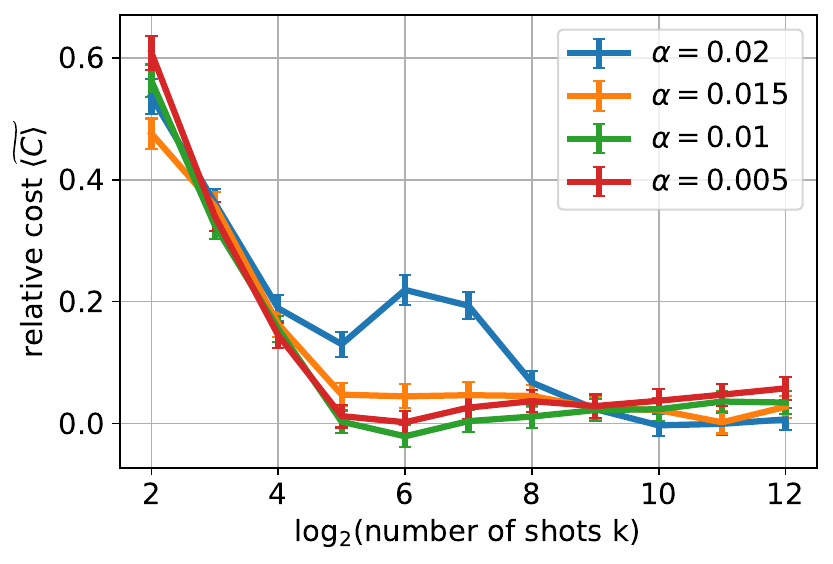}
    \caption{QuEnc efficiency on {\tt MaxCut}-16 problem with different numbers of shots and learning rates relative to $k=\infty$. X-axis: number of shots in log-scale, y-axis: relative cost function Eq.~\eqref{eq:rel_cost}. 1.0-level corresponds to learning with $k=\infty$ and $\alpha=\alpha_0(16)=0.02$. All curves were obtained by averaging over 250 problems.}
    \label{fig:shots1}
\end{figure}

For each problem size, circuit depth, and the number of shots $k$, the learning rate needs to be tuned.
As a representative example of this, we present the dependence of the relative cost function on the number of shots for different learning rates in the context of {\tt MaxCut} with $n_c=16$ nodes (see Fig.~\ref{fig:shots1}).
As can be seen, it is optimal to perform $k = 2^6$ measurements with $\alpha=\alpha_0/2$ to reach efficiency no less than with $k\to\infty$, where $\alpha_0=0.02$ is the number of shots that was adjusted for $k\to\infty$ case.

\subsection{Warm-start optimization} 
\label{AppendixSecWarm}

There are several techniques that can be used to improve the performance of the optimization algorithm: one of these is warm-start optimization. 
The solutions of continuous-valued relaxation can be used to initialize hybrid algorithms, which is known as warm-starting them~\cite{Gondzio1998}, and the warm-start of the QAOA was recently introduced~\cite{Egger2021} and analyzed~\cite{Truger2022}.

The structure of the QuEnc also allows for wise initialization: starting from a certain quantum state, we improve it by adjusting variational gates.
Firstly, we encode a binary string into the quantum state defined in Eq.~\eqref{eq:quenc_quantum_state}.
Let us define $\beta_i = \frac{1}{\sqrt{n_c}}$
and $a_i$ and $b_i$ as 0 or 1, corresponding to the value of $x_i$.
For instance, in the case of the starting point $\vec{x} = [0, 1, 1, 0 ]$, the state is prepared as:
\begin{equation}
	\ket{\psi_{init}} = \frac{1}{2} \left( \ket{0}_a \ket{0}_r + \ket{1}_a \ket{1}_r + \ket{1}_a \ket{2}_r  + \ket{0}_a \ket{3}_r \right).
\end{equation}

\begin{figure}[ht]
	\noindent\centering{
		\includegraphics[scale=0.25]{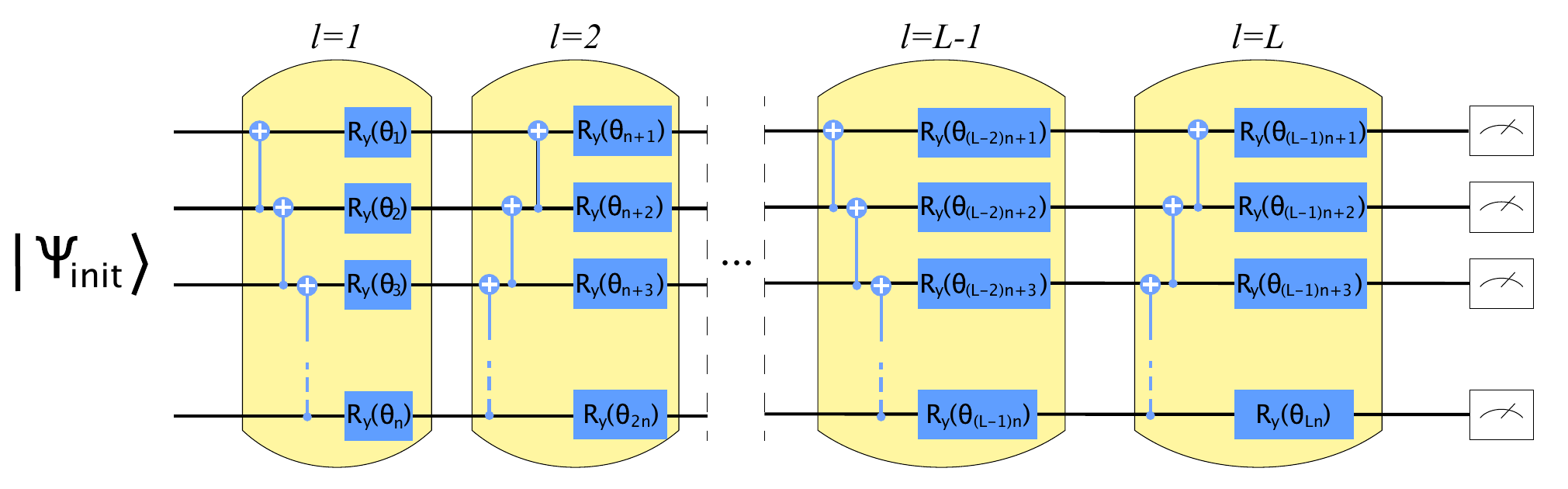}
	}
	\caption{
 The warm-start protocol of the QuEnc. 
		Zero $\vec{\theta}$ doesn't change the initial state $\ket{\psi_{init}}$. 
        Optimization technique of the circuit parameters improves the initial solution.
		Here, $n = \log(n_c)$ is the number of qubits, $n_c$ is number of classical variables and $L$ is even number of layers.}
	\label{fig:end_quenc}
\end{figure}

Secondly, we tune this state in order to improve the objective values.
In order to gradually change the state, we modify the ansatz so that the transformation is the identity if $\vec{\theta}$ are initialized as zeros.
The changed quantum circuit is shown in Fig.~\ref{fig:end_quenc}, where we exclude Hadamard gates, use only an even number of layers, and place the CNOTs in reverse order for the closest layers.
Zero $\vec{\theta}$ doesn't change the starting point but by perturbing the values in order to escape local minima, gradient descent or other optimization techniques are used to improve the solution.

\section{Constrained Optimization}

\subsection{Proof of commuting of projections}
\label{AppendixSecCommuting}

Recall that every classical bit corresponds to a pair (ancilla in 0 and 1) of components of the state vector of QuEnc $\ket{\psi}\in\mathbb{R}^{2n_c}$. 
Let's consider the case with 2 constraints: $x_0+x_1=1$ and $x_2+x_3=1$.
Considering the above and the fact that projection must not act on unconstrained bits, their corresponding projections have the following form: 
\begin{equation}
\begin{aligned}
    P_1 &=
\begin{pmatrix}
        V &  & O \\
         & Id &  \\
        O &  & Id \\
    \end{pmatrix},
    P_2=
\begin{pmatrix}
        Id &  & O \\
         & V &  \\
        O &  & Id \\
    \end{pmatrix},
\end{aligned}
\end{equation}
where $Id$ is the identity matrix.
It is obvious that $P_1P_2-P_2P_1=0$. 

The case with constraints $x_i+x_j=1$ and $x_k+x_l=1$, where $i,j,k,l$ are pairwise different, is derived from the previous one by permutation of the basis vectors. The matrices of the projections could be written as $P'_1=S^{-1}P_1 S$, $P'_2=S^{-1}P_2 S$, where $S$ is a transition matrix. They also commute as $P'_1P'_2-P'_2P'_1=S^{-1}P_1 SS^{-1}P_2 S-S^{-1}P_2 SS^{-1}P_1 S=S^{-1}(P_1P_2-P_2P_1)S=0$.

\section{Analysis}

\subsection{Tensor Networks contraction algorithm}
\label{AppendixSecContraction}

Tensor network contraction can be represented as a sequence of pairwise tensor contractions. For a given tensor network contraction path, one can quantify the space and time costs. First, the total space required for the contraction of a tensor network is given 
by
\begin{equation}
    S = \max\limits_T |dim(T)|
\end{equation}
taken over all tensors $T$ formed in the contraction process.
Similarly, the time complexity of the contraction path can be defined as
\begin{equation}
    C = \sum\limits_{T = A \times B} |dim(A)|\cdot |dim(B)|/|dim(A\cap B)|, 
\end{equation} 
where $|dim(A\cap B)|$ is the total dimension of the indices on which the contraction of $A$ and $B$ into $T$ takes place (for all tensors $T$ formed in the contraction process). Thus, one can find a \textit{space-optimal} or a \textit{ time-optimal contraction path} for a given TN by minimization $S$ or $C$, respectively. 

There are many ways to find a sub-optimal contraction path. For example, the Hyper-Greedy approach~\cite{Cotengra}: on each step, heuristically score each possible pairwise contraction $T_i \times T_j = T_k$ using a cost function $cost(T_i,T_j) = size(T_k) - \alpha(size(T_i)+size(T_j))$ and choose this pair of tensors to contract with probability $p(T_i,T_j) \propto \exp{(-cost(T_i,T_j)/\tau)}$ where $\alpha$ and $\tau$ are tunable constants. 
Hyper-Greedy generally outperforms other greedy approaches and is quick to run, making it a simple but useful reference algorithm. This approach is not guaranteed to find the global minimum of $S$ or $C$, yet by simply sampling many paths and tuning the heuristic parameters, one can often get arbitrarily close to the optimum.

\subsection{Expressibility}
\label{AppendixSecExpressibility}

We evaluate the expressibility through the fidelity $F=\left|\left\langle\psi_{\theta} \mid \psi_{\phi}\right\rangle\right|^{2}$, where parameters $\theta \in \Theta$ and $\phi \in \Phi$ are randomly sampled from the circuit ansatz $C$.
Here, $P_{Haar}(F)$ is the probability distribution
of the fidelity $F=\left|\left\langle\psi \mid \psi^{\prime}\right\rangle\right|^{2}$, where $|\psi\rangle$ and $\left|\psi^{\prime}\right\rangle$ are sampled according to the Haar measure~\cite{sim2019expressibility}. 
We quantify the circuit ansatz $C$ expressibility as
\begin{equation}
    \mathcal{E}(\mathrm{C})=\int_{0}^{1} P(C, F) \log \frac{P(C, F)}{P_{\text {Haar}}(F)} d F.
    \label{eq:q_expr}
\end{equation}

We compare the Sequential-2QG ansatz from Fig.~\ref{qcn} and the Simultaneous-2QG ansatz with alternating $R_y$ and $R_z$ gates, see Fig.~\ref{fig:step_circuit}(a). 

\begin{figure}[h!]
    \centering
    \includegraphics[scale=0.55]{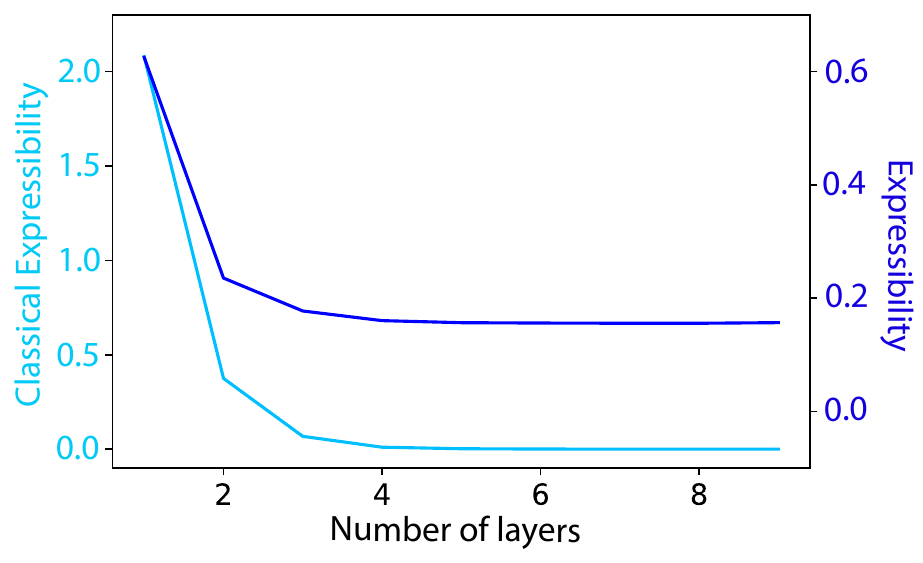}
    \caption{Expressibility from Ref.~\cite{sim2019expressibility} and expressibility based on classic solutions of QuEnc ansatz.}

    \label{fig:classic}
\end{figure}

We observe that classical expressibility is highly correlated with the one introduced for quantum states.
During the problem solving, we are not interested in complete and uniform coverage of all states but in covering all solutions. 
In this regard, to calculate the representative power, you can compare the uniform distribution with the solutions at the output of the circuit using the Kullback–Leibler divergence:
\begin{equation}
    \mathcal{E}(\mathrm{C})=\int_{0}^{1} P(C, S) \log \frac{P(C, S)}{P_{\text {Uniform}}(S)} d S,
\end{equation}
where $P(C, S)$ is probability distribution of the solutions $S$ at the output of the circuit and $P_{\text{Uniform}}$ is the uniform distribution that we strive for.
Both expressibilities for the Sequential-2QG circuit with three qubits are shown in Fig.~\ref{fig:classic}.

\bibliography{aipsamp.bib}

\end{document}